\documentclass[usenatbib]{mnras}

\usepackage[T1]{fontenc}

\DeclareRobustCommand{\VAN}[3]{#2}
\let\VANthebibliography\thebibliography
\def\thebibliography{\DeclareRobustCommand{\VAN}[3]{##3}\VANthebibliography}

\usepackage{graphicx}	
\graphicspath{{Images_HR/}}
\usepackage{amsmath}	
\usepackage{amssymb}	
\usepackage[table,xcdraw]{xcolor}    
\usepackage{subfig}
\usepackage{comment}
\usepackage[normalem]{ulem}
\usepackage{physics}
\usepackage{xcolor}
\usepackage{enumitem}

\definecolor{royalazure}{rgb}{0.0, 0.22, 0.66}
\definecolor{auburn}{rgb}{0.43, 0.21, 0.1}
\definecolor{bostonuniversityred}{rgb}{0.8, 0.0, 0.0}
\definecolor{planet}{HTML}{69DF45}

\usepackage{hyperref}
\hypersetup{
    colorlinks=true,
    linkcolor=bostonuniversityred,
    linktoc=all,
    filecolor=blue,      
    urlcolor=royalazure,
    citecolor=royalazure
}

\usepackage{lipsum}

\usepackage{newtxtext,newtxmath}

\title[Dust Dynamics in a GI Disc]{The Role of Drag and Gravity on Dust Concentration in a Gravitationally Unstable Disc}
\author[Rowther et al.]{Sahl Rowther,$^{1, 2, 3}$\thanks{E-mail: sahl.rowther@leicester.ac.uk}
Rebecca Nealon,$^{1,2}$
Farzana Meru$^{1,2}$,
James Wurster$^{4}$,
\newauthor
Hossam Aly$^{5,6}$,
Richard Alexander$^3$, 
Ken Rice$^{7,8}$,
Richard A. Booth$^{9}$
\\
$^{1}$Centre for Exoplanets and Habitability, University of Warwick, Coventry CV4 7AL, UK\\
$^{2}$ Department of Physics, University of Warwick, Coventry CV4 7AL, UK\\
$^{3}$ School of Physics and Astronomy, University of Leicester, Leicester LE1 7RH, UK\\
$^{4}$Scottish Universities Physics Alliance (SUPA), School of Physics and Astronomy, University of St. Andrews, North Haugh, St Andrews, Fife KY16 9SS, UK \\
$^{5}$  Faculty of Aerospace Engineering, Delft University of Technology, Kluyverweg 1, 2629 HS Delft, The Netherlands \\
${6}$ Zentrum für Astronomie der Universität Heidelberg, Astronomisches Rechen-Institut, Mönchhofstr. 12-14, 69120 Heidelberg, Germany \\
$^{7}$ SUPA, Institute for Astronomy, University of Edinburgh, Royal Observatory, Blackford Hill, Edinburgh EH9 2HJ, UK \\
$^{8}$ Centre for Exoplanet Science, University of Edinburgh, Edinburgh, EH9 3FD, UK\\
$^{9}$ School of Physics and Astronomy, University of Leeds, Leeds, LS2 9JT, UK
}

\date{Accepted XXX. Received YYY; in original form ZZZ}

\pubyear{2022}

\begin{document}
\label{firstpage}
\pagerange{\pageref{firstpage}--\pageref{lastpage}}
\maketitle

\begin{abstract}
    We carry out three dimensional smoothed particle hydrodynamics simulations to study the role of gravitational and drag forces on the concentration of large dust grains $(\text{St} > 1)$ in the spiral arms of gravitationally unstable protoplanetary discs, and the resulting implications for planet formation. We find that both drag and gravity play an important role in the evolution of large dust grains. If we include both, grains that would otherwise be partially decoupled will become well coupled and trace the spirals. For the dust grains most influenced by drag (with Stokes numbers near unity), the dust disc quickly becomes gravitationally unstable and rapidly forms clumps with masses between $0.15 - 6M_\oplus$. A large fraction of clumps are below the threshold where runaway gas accretion can occur. However, if dust self-gravity is neglected, the dust is unable to form clumps, despite still becoming trapped in the gas spirals. When large dust grains are unable to feel either gas gravity or drag, the dust is unable to trace the gas spirals. Hence, full physics is needed to properly simulate dust in gravitationally unstable discs.
    Dust trapping of large grains in spiral arms of discs stable to gas fragmentation could explain planet formation in very young discs by a population of planetesimals formed due to the combined roles of drag and gravity in the earliest stages of a disc's evolution. Furthermore, it highlights that gravitationally unstable discs are not just important for forming gas giants quickly, it can also rapidly form Earth mass bodies.
\end{abstract}




\section{Introduction}

In recent years, the Atacama Large Millimeter/submillimeter Array (ALMA) has revealed that many of the brightest, and presumably the most massive, protoplanetary discs are highly structured. The most common of these sub-structures are rings \& gaps \citep{2015ALMA,2016Andrews,2018Andrews,2018Fedele,2018Huang,2018Dipierro,2018Long,2020Booth}. A natural explanation for the origin of ring \& gap structures is through planet-disc interaction \citep[see reviews by][]{2012Kley,2022Sijme}. The evidence of planets being the cause has been strengthened through studies of the gas kinematics \citep{2015Perez,2018bPinte,2019Pinte,2020Pinte,2021Calcino,2023Pinte}. 

The discs with observed ring \& gap structure are young, with some being less than 1Myr old \citep{2015ALMA,2018Fedele,2018Dipierro,2018Sheehan,Segura-Cox2020}. Forming massive planets at these locations is challenging at such young ages.  Many of the potential planets inferred from the rings \& gaps and kinematics are massive ($0.1-10~M_\mathrm{Jup}$) and far from the central star at tens or even hundreds of \textsc{au} \citep{2019Lodato,2022Pinte}.   Traditional planet formation models such as core accretion struggle to form massive planets at wide orbital separations so quickly. Hence, a natural question to ask is how these putative planets could have formed so quickly.

An alternate planet formation mechanism is through gravitational instability. In the earliest stages of a disc's evolution, it is expected to be massive enough to develop gravitational instabilities in the form of spiral structures. Although this phase is short-lived, there is observational evidence of such discs \citep{2016Perez,2018bHuang,2021Paneque}, which is also supported by theoretical predictions \citep{2017Meru,2020Hall,2021Longarini}. In this scenario, if the disc is massive enough, it can fragment and form giant planets \citep{1997Boss} on dynamical timescales \citep{2001Gammie}. An issue is that although the initial mass of these fragments is in the planetary regime, they quickly accrete a lot of mass and become brown dwarfs ($\,{>}13M_\mathrm{Jup}$) as they migrate \citep{2018Stamatellos}. However, there is another path to forming planets if the disc is stable to fragmentation, but still massive enough to form spirals. The spirals are regions of high pressure where dust can become trapped and grow to form planetesimals which become the seeds of planet formation \citep{2004Rice,2006Rice,2012Gibbons,2016Booth,2020Elbakyan,2021Baehr,2022Baehr,2023Longarini,2023bLongarini}. On the other hand, it has also been shown that gravitational instability can inhibit the concentration of dust \citep{2013Walmswell,2020Riols}. The differing conclusions could be due to the widely different methodologies which range from 2D local shearing boxes to 3D global simulations, and from using dust particles in the test particle regime to feeling gravitational acceleration from both gas and itself. Thus, the aims of this study (using 3D global simulations) are to determine what drives the dynamics of large dust grains in these spirals; the drag force, or gravity? 

In this study we perform three-dimensional global numerical simulations to investigate the role of drag and gravity on the dust dynamics in a gravitationally unstable protoplanetary disc. In \S\ref{sec:model} we describe the simulations presented in this study. In \S\ref{sec:results} we present the results of dust concentration in the spiral arms, and roles of drag and gravity. The limitations and implications of our work are discussed in \S\ref{sec:Discussion}.

\section{Model}
\label{sec:model}

We use \textsc{Phantom}, a smoothed particle hydrodynamics (SPH) code developed by \cite{2018Price} to perform the suite of simulations presented here. SPH codes are favourable for scenarios with high density contrasts. Thus, \textsc{Phantom} is well suited for the problem at hand and has been previously used for dusty simulations \citep{2012aLaibe,2012bLaibe,2015Dipierro,2020Price} and for simulations involving self-gravity \citep{2020Rowther,2020Cadman,2020bRowther,2022Rowther,2022bRowther}. For the simulations in this work, the dust and gas are modelled as two separate fluids \citep{2012aLaibe,2012bLaibe}.

\subsection{Disc setup} 
All discs are modelled using $2 \times 10^6$ gas particles and $2.5 \times 10^5$ dust particles between $R_\mathrm{in} = 4$\textsc{au}  and $R_\mathrm{out} = 100$\textsc{au}  where the fiducial disc has a mass of $0.2M_{\odot}$ disc around a $1M_{\odot}$ star. A sink particle \citep{1995Bate} is used to model the central star. The accretion radius of the central star is set to be equal to the disc's inner boundary, $R_\mathrm{in}$. Particles within 0.8 times the accretion radius are accreted immediately without any checks. The surface density profile $\Sigma$ is given by
\begin{equation}
\Sigma = \Sigma_{0}  \left ( \frac{R}{R_{\mathrm{in}}} \right)^{-1} f_{s},
\end{equation}
where $\Sigma_{0} = 1.1 \times 10^3\mathrm{g\ cm}^{-2}$,  and ${f_{s} = 1-\sqrt{R_\mathrm{in}/R}}$ is {the factor used to smooth the surface density at the inner boundary of the disc}. The initial temperature profile is expressed as a power law
\begin{equation}
T = T_{0} \left ( \frac{R}{R_{0}} \right)^{-0.5},
\end{equation}
where $T_{0}$ is set such that the disc aspect ratio ${H/R=0.05}$ at ${R=R_{0}=R_\mathrm{in}}$. The energy equation for the gas particles is 
\begin{equation}
\label{eq:eng}
\frac{\mathrm{d} u_\mathrm{g}}{\mathrm{d} t} = -\frac{P_\mathrm{g}}{\rho_\mathrm{g}} \left ( \nabla \cdot \vb*{v}_\mathrm{g} \right) + \Lambda_{\mathrm{shock}} - \frac{\Lambda_{\mathrm{cool}}}{\rho_\mathrm{g}}
\end{equation}
where we assume an adiabatic equation of state with $\gamma = 5/3$, and $u$ is the specific internal energy, $P$ is the pressure, $\rho$ is the density and $\vb*{v}$ is the velocity. The subscript `$\mathrm{g}$' refers to gas particles. We do not include drag heating from back-reaction of the dust on the gas. The first term on the RHS is the $P\mathrm{d}V$ work, and $\Lambda_{\mathrm{shock}}$ is a heating term that is due to the artificial viscosity used to correctly deal with shock fronts. The final term  
\begin{equation}
\Lambda_{\mathrm{cool}} = \frac{\rho_\mathrm{g} u_\mathrm{g}}{t_{\mathrm{cool}}}
\end{equation}
controls the cooling rate in the disc. Here the cooling time is straightforwardly implemented to be proportional to the dynamical time by a factor of $\beta$,
\begin{equation}
t_{\mathrm{cool}} = \beta \Omega^{-1}
\end{equation}
where $\Omega$ is the orbital frequency, and $\beta=15$. 
The \cite{2010Cullen} switch detects any shocks that form and generates the correct artificial viscosity response. The linear artificial viscosity parameter $\alpha_{\text{AV}}$ varies depending on the proximity to a shock. Close to the shock, it takes a maximum of $\alpha_{\mathrm{max}} = 1$, and a minimum of ${\alpha_{\mathrm{min}} = 0.1}$ far away. The quadratic artificial viscosity coefficient $\beta_{\text{AV}}$ is set to 2 \cite[see][]{2018Price,2015Nealon}.

\subsection{Dust dynamics}
\label{sec:DD}

The dust dynamics is governed by gravitational and drag forces from interactions with the gas. The acceleration of the dust particles is given by 
\begin{equation}
    \label{eq:DD}
    \frac{\mathrm{d} \vb*{v}_\mathrm{d}}{\mathrm{d} t} = -\frac{K}{\rho_\mathrm{d}} (\vb*{v}_\mathrm{d} - \vb*{v}_\mathrm{g}) + \vb*{a}_\mathrm{dust-gas} + \vb*{a}_\mathrm{dust-dust} + \vb*{a}_\mathrm{dust-star},
\end{equation}
where the subscript `$\mathrm{d}$' refers to dust particles. The first term on the right hand side is the drag force which acts to eliminate any velocity difference between the dust and gas. The second term on the right hand side of Equation \ref{eq:DD} is the gravitational acceleration due to interactions between dust and gas particles; the third term is the dust self-gravity; and the final term is the gravitational acceleration from the central star. 

An important timescale for the drag force is the stopping time, $t_\mathrm{s}$, the timescale at which the differential velocity between the dust and gas decays. If the stopping time is short, the dust velocities quickly decay to the gas velocities resulting in the dust becoming coupled to the gas. If it's long, the dust velocities will remain different to the gas velocities, and hence the dust structure will be different from the gas structure. The stopping time is related to the drag coefficient $K$ \citep{2012bLaibe}, by
\begin{equation}
    t_\mathrm{s} = \frac{\rho_\mathrm{g} \rho_\mathrm{d}}{K\rho},
\end{equation}
where $\rho = \rho_\mathrm{g} + \rho_\mathrm{d}$. However, to decrease computational expense when dust clumps form,
we instead approximate $\rho$ to be equal to $\rho_\mathrm{g}$ only when determining the minimum timestep required for evolving the particle. The effectiveness of the drag force can be described by the Stokes number St, which is defined by the ratio of the stopping time to the orbital timescale,
\begin{equation}
    \mathrm{St} = t_\mathrm{s} \Omega.
\end{equation}
The simulations in this study fall in the Epstein drag regime where the stopping time is approximated by 
\begin{equation}
    t_\mathrm{s} = \frac{\rho_\text{grain} s_\text{grain}}{\rho c_\text{s}f} \sqrt{\frac{\pi \gamma}{8}},
\end{equation} where $\gamma = 5/3$ is the adiabatic index, $c_\text{s}$ is the sound speed, $s_\text{grain}$ is the grain size, \smash{$\rho_\text{grain} = 3 \text{g/cm}^3$} is the intrinsic grain density, and $f$ is a correction for supersonic drift velocities given by \cite{1975Kwok} as 
\begin{equation}
f = \sqrt{1 + \frac{9\pi}{128} \frac{\Delta v^2}{c_s^2}},
\end{equation}
where $\Delta_v \equiv | \vb*{v}_d - \vb*{v}_g |$.  For small Stokes numbers (${<}\,1$), the stopping time is sub-orbital, and the drag force quickly eliminates any velocity difference between the dust and gas. However, for large Stokes numbers ($>1$), it can take multiple orbits before the velocity difference decays. Hence for a dynamic disc where the substructure continuously evolves -- as is the case for a gravitationally unstable disc -- it becomes difficult for the dust velocities to decay to the gas velocities.

\subsection{Defining Toomre Q for dust}

The Toomre Q parameter \citep{1964Toomre} gives a measure of how gravitationally unstable a disc is, and is defined as 
\begin{equation}
    \label{eq:Toomre}
    Q_\mathrm = \frac{c_{\text{s}} \Omega}{\pi G \Sigma},
\end{equation}
where $G$ is the gravitational constant, $c_\text{s}$ is the sound speed, and $\Sigma$ is the surface density. 
A disc becomes more gravitationally unstable as the surface density increases, or as the sound speed decreases. Although the above equation is generally used for the gas, we calculate the Toomre Q of dust using the effective sound speed and surface density of the dust. The effective sound speed of the dust particles $c_{\text{s}}^\mathrm{dust}$ can be obtained by calculating their velocity dispersion as
\begin{equation}
    \left(c_{\text{s}}^\mathrm{dust}\right)^2_i = \sum_{j=1}^{N_\mathrm{neigh}} m_j \frac{(v_{\mathrm{d},i} - v_{\mathrm{d},j})^2}{\rho_{\mathrm{d},j}} W_{ij}(h_i),
\end{equation}
where the calculation is summed over all dust particles that are neighbours of particle $i$. The mass and density are given by $m_j$ and $\rho_{\mathrm{d},j}$, and $v_{\mathrm{d}, i}$ and $v_{\mathrm{d}, j}$ are the magnitudes of the velocity of dust particles $i$ and $j$, respectively. The smoothing kernel and length are given by $W_{ij}$ and $h_i$, respectively.

\begin{figure*}
    \centering
    \includegraphics[width=\linewidth]{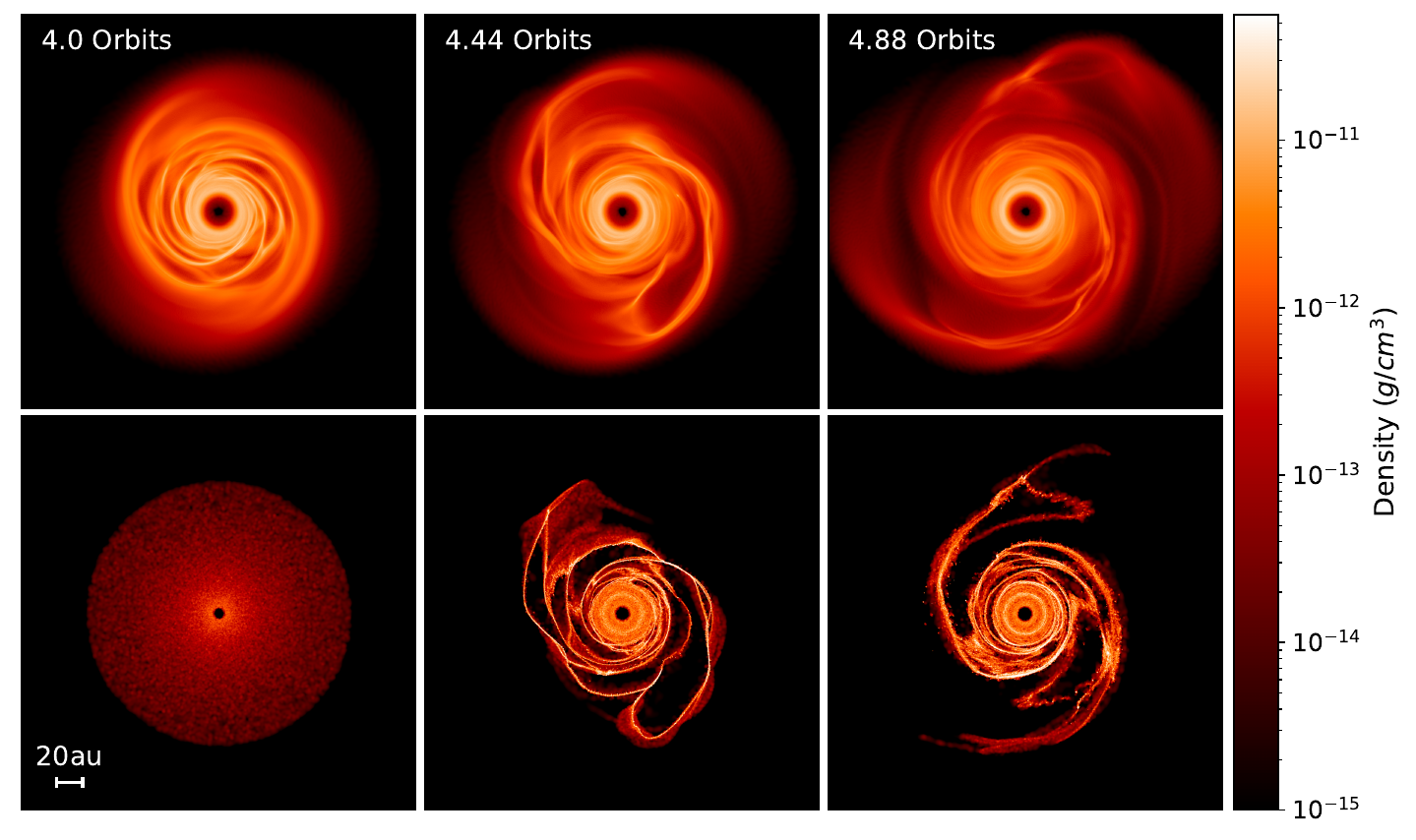}
    \caption{Cross section slices of density in the $z=0$ plane of gas, i.e. the midplane, (top) and 50cm sized dust grains (bottom) at $t=4, 4.44,$ and $4.88$ outer orbits (from left to right). The dust is initialised at $ t=4$ orbits after spiral structures have formed in the gas. Dust rapidly concentrates in the spirals becoming gravitationally unstable and forming clumps as seen at $t=4.44,$ and $4.88$ outer orbits. At $t=4.88$ outer orbits, a few of the dust clumps have become massive enough to cause spiral wakes which are visible in both the gas and dust.}
    \label{fig:50cm}
\end{figure*}

\begin{figure}
    \centering
    \includegraphics[width=0.99\linewidth]{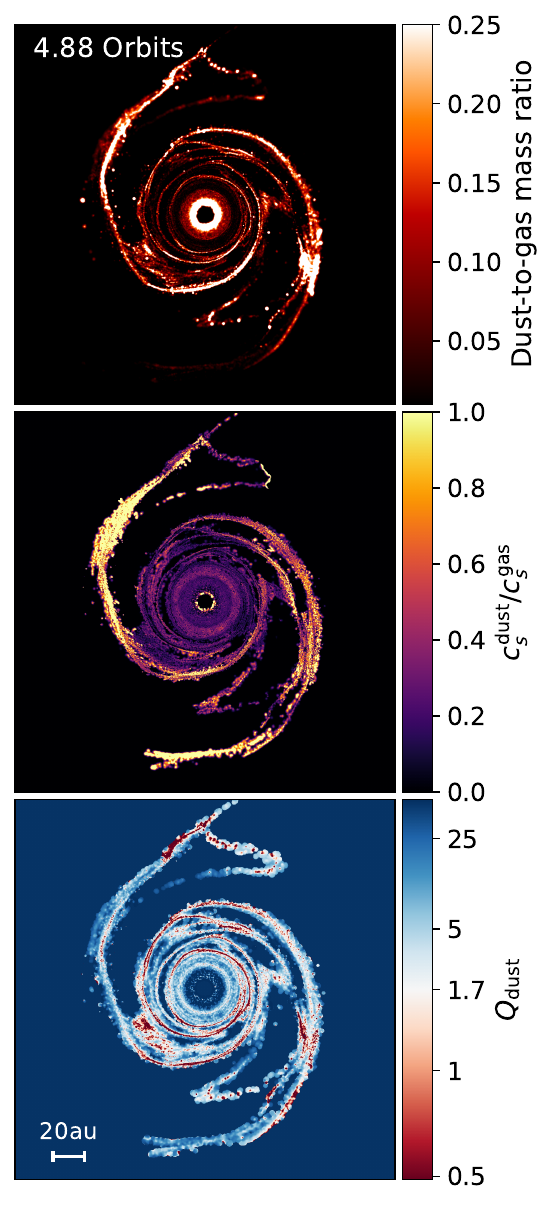}
    \caption{The top and middle panels show the cross section slices of the dust-to-gas mass ratio and the sound speed of dust relative to the gas, respectively in the $z=0$ plane. The bottom panel shows Toomre $Q$ of the dust disc. The combined effect of the increased dust enhancement in the spirals and lower sound speed results in the dust disc becoming locally unstable in the spiral arms (red regions).}
    \label{fig:Q_plot}
\end{figure}

\begin{figure}
    \centering
    \includegraphics[width=\linewidth]{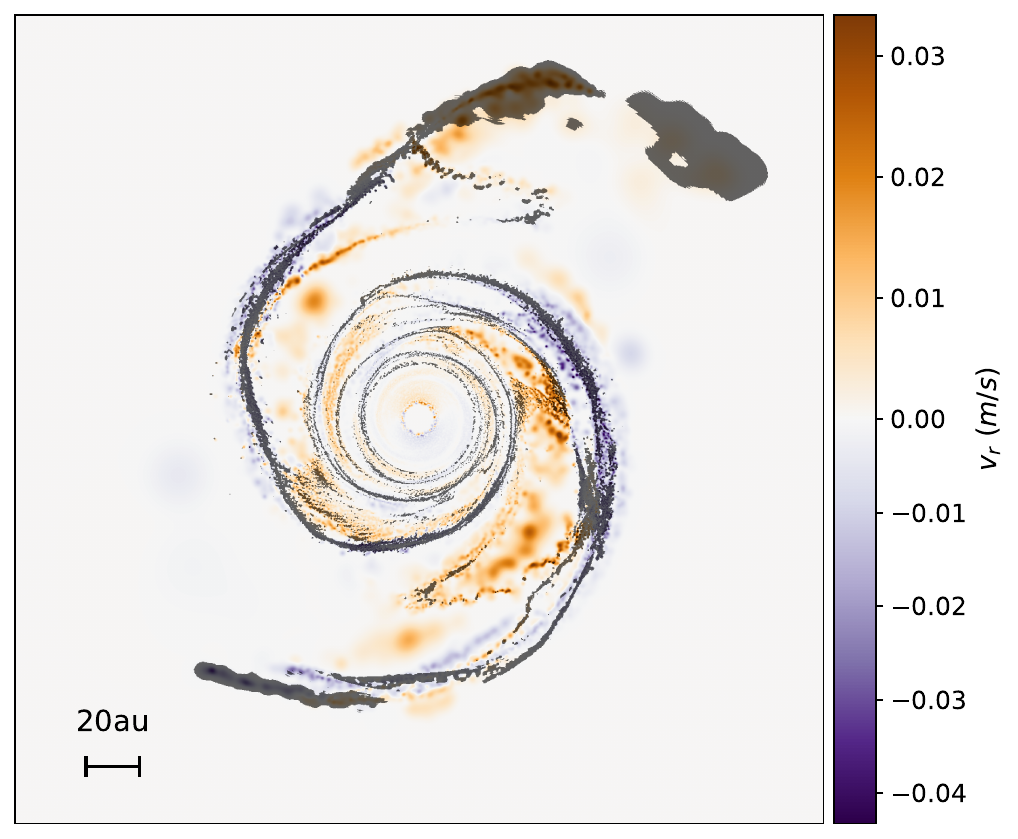}
    \caption{Cross section slice of the radial velocity $v_r$ of the dust in the $z=0$ plane. The dust drifts towards the gas spirals which results in dust becoming trapped. The presence of gravitational instabilities could slow down radial drift of dust towards the central star. The orange and purple regions correspond to outward and inward dust motion, respectively. We can clearly see drift towards the spiral arms from both directions.}
    \label{fig:vr}
\end{figure}

\begin{figure*}
    \centering
    \includegraphics[width=1\linewidth]{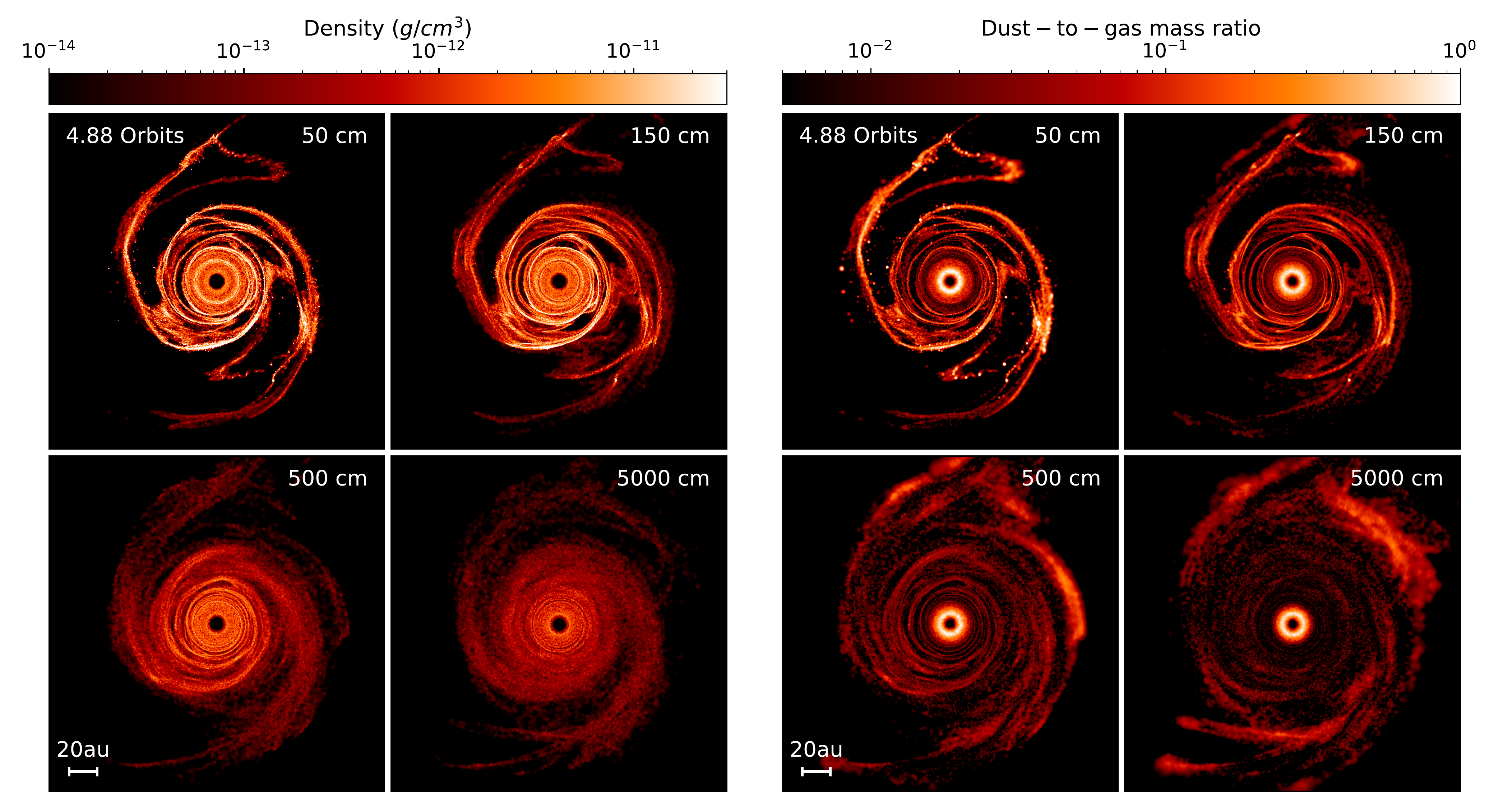}
    \caption{Cross section slice of the dust density (left panels) and the dust-to-gas mass ratio (right panels) in the $z=0$ plane 1.5 outer obits after the dust was added. Each panel represents a different size: 50\,cm (top left), 150\,cm (top right), 500\,cm (bottom left), and 5000\,cm (bottom right). The corresponding Stokes numbers are roughly 4, 12, 40, and 400 respectively. As the dust size increases, dust trapping becomes less efficient due to the decreasing influence of the drag force. Only the smallest two dust sizes become gravitationally unstable and form clumps.}
    \label{fig:allStokes}
\end{figure*}

\begin{figure*}
    \centering
    \includegraphics[width=\linewidth]{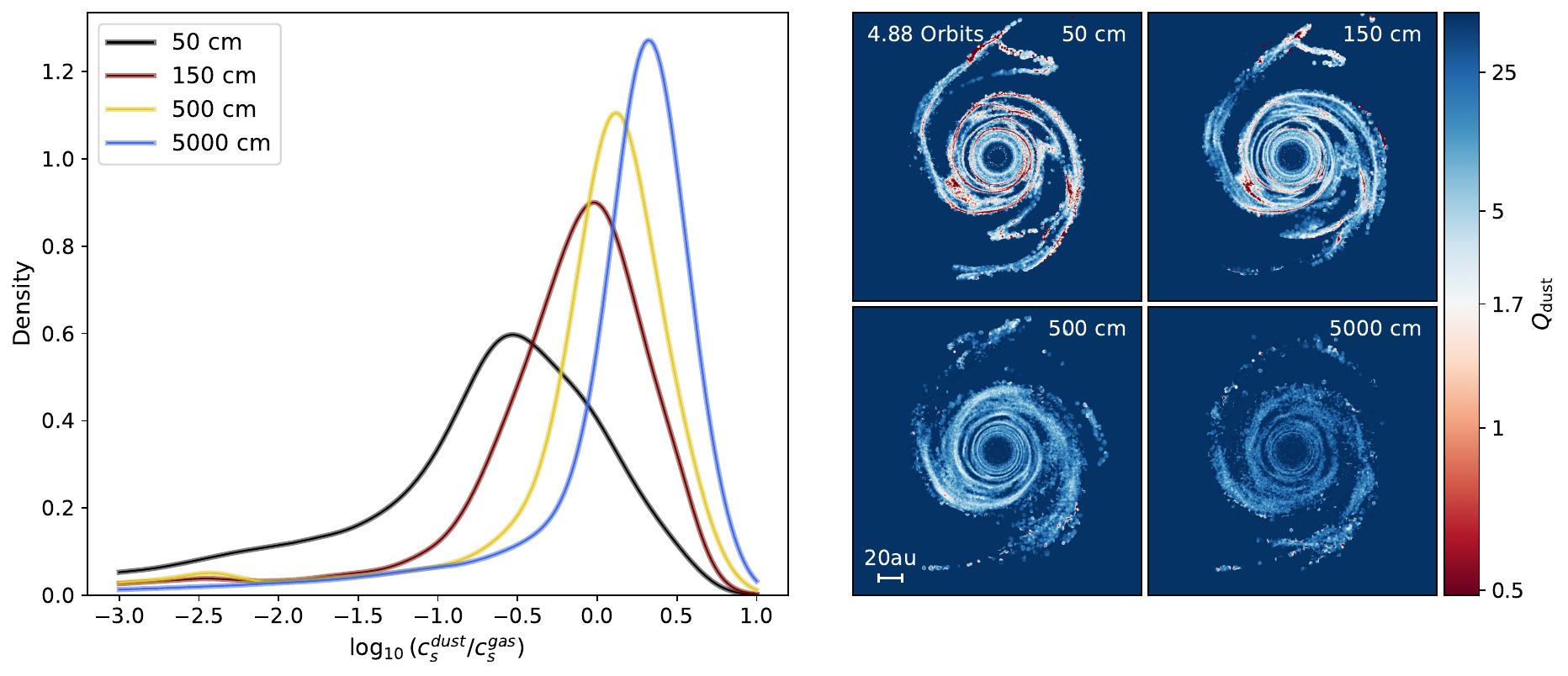}
    \caption{A probability density function (PDF) of the sound speed of dust relative to the gas, $c_\text{s}^{\text{dust}}/c_\text{s}^{\text{gas}}$ (left panel), for all four dust sizes as described in Figure \ref{fig:allStokes}. The drag force becomes less effective with increasing dust size. Hence, the dust motion becomes uncorrelated with the gas as seen by the increasing velocity dispersion with increasing dust size. The right panels show the Toomre Q of the dust disc for all four dust sizes. The less efficient dust trapping with increasing dust size results in the dust becoming more stable.}
    \label{fig:csStokes}
\end{figure*}

\begin{figure*}
    \centering
    \includegraphics[width=0.95\linewidth]{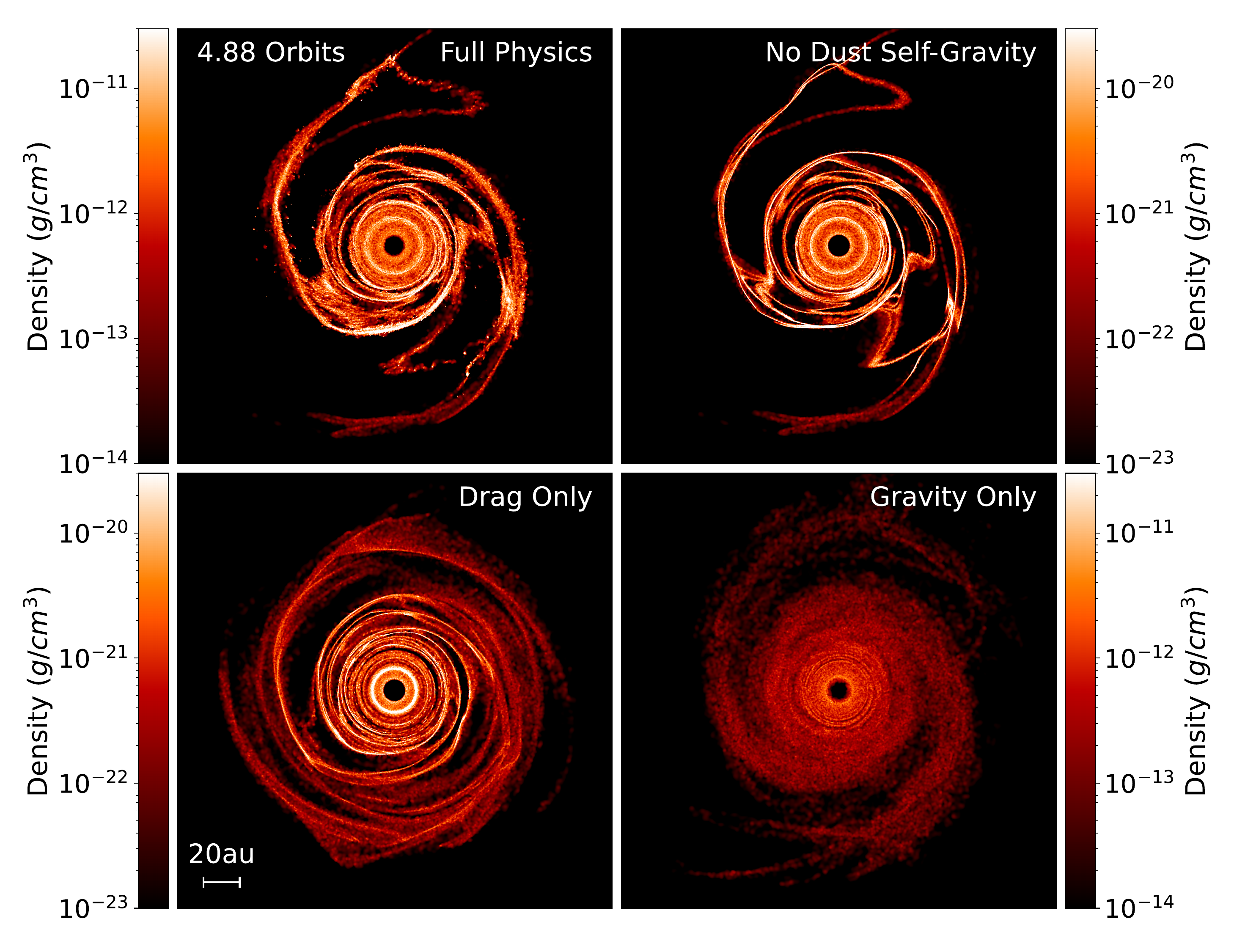}
    \caption{Cross section slices of dust density in the $z=0$ plane 0.88 outer orbit after physics has modified. The four simulations are: the reference simulation with physics as normal (Full Physics - top left panel), the dust no longer feels the gravitational force of other dust particles (No Dust Self-Gravity - top right panel), the dust no longer feels gravitational acceleration from either the gas or dust (Drag Only - bottom left panel), and the dust no longer feels the drag force (Gravity Only - bottom right panel). The dust densities are lower in the top right and bottom left panels because the dust particle mass is decreased by eight orders of magnitude to mimic turning off dust self-gravity. The loss of dust self-gravity results in clumps being unable to form despite  the dust still becoming trapped in the gas spirals. Neither the drag force nor gravity is able to couple the dust to the gas. Both are necessary for the dust to trace the gas spirals. See the top right panel in Figure \ref{fig:50cm} for the gas distribution. The spiral structures visible in the bottom panels eventually disappear in the absence of drag/gravity (see Figure \ref{fig:longTerm}).}
    \label{fig:allPhysics}
\end{figure*}

\subsection{The suite of simulations}
\label{sec:suite}

Initially, the disc is modelled as a gas only simulation until spiral structures have developed. After 4 outer orbits, the dust is added uniformly as a separate dust disc as if it is $t=0$ where no spirals are present (see bottom left panel of Figure \ref{fig:50cm}). In the fiducial simulation, 50\,cm sized dust grains are used. This corresponds to an initial average Stokes number of 4 at $R=50$\textsc{au}.

The effectiveness of the drag force is dependent on $t_\mathrm{s}$, and thus the size of the dust grains, where the larger the dust grain, the less efficient the drag force becomes. Since the drag force acts  to eliminate any velocity difference between the dust and gas, a weakening drag force leads to decoupling of the dust.  Hence, to determine when dust should stop tracing the gas spirals in a gravitationally unstable disc, three additional simulations were modelled with dust sizes of 150, 500, and 5000\,cm. This corresponds to initial average Stokes numbers of 12, 40, and 400, respectively.

To investigate the importance of gravitational and drag forces, the simulation with 50\,cm sized grains at $t=4$ outer orbits is used as the initial condition for four further simulations. The initial conditions can be seen in the left column of Figure \ref{fig:50cm}. Except for the reference simulation, each simulation turns off certain physics to see how the dust evolves in its absence. The four simulations are 
\begin{description}[itemsep=2pt,parsep=4pt]
    \item[Full Physics --] The reference simulation. The evolution of the fiducial disc is continued as normal where the dust feels both the gravity of gas and dust, as well as the drag force.
    \item[No Dust Self-Gravity --] The dust no longer feels the gravity of other dust particles. It only feels the gas gravity and the drag force.  Computationally, dust self-gravity is turned off by decreasing the dust particle mass by eight orders of magnitude. This makes the third term in Eq \ref{eq:DD}, $\vb*{a}_\mathrm{dust-dust}$ negligible but does not affect the other terms.
    \item[Drag Only --] The dust is no longer affected by gravity. It only feels the drag force. The second term in Eq \ref{eq:DD}, $\vb*{a}_\mathrm{dust-gas}$ is set to 0, in addition to the decreased dust particle mass. However, since the gas still feels its own self-gravity, the velocity of the gas particles is increased due to the gas disc itself. If the dust has no knowledge of the gas disc at all, the dust particles will orbit at a slower velocity relative to the gas, resulting in outward migration. Thus, the dust velocity includes a correction for the enclosed mass in the Keplerian velocity. The dust therefore has no local knowledge of any gravitational potential, but only the global gas disc.
    \item[Gravity Only --] The drag force has no effect on the dust. Only gas and dust gravity acts on the dust. Here, only the first term (the drag force) in Eq \ref{eq:DD} is set to 0. This is a similar scenario to the simulations in \cite{2013Walmswell}.
\end{description}
In all simulations, the dust always feels the gravitational acceleration from the central star (the final term in Eq \ref{eq:DD}), and the gas always feels its own self-gravity.

\subsection{Clump finding}
\label{sec:clumps}

We use the clump finding algorithm introduced in \cite{2023Wurster}. In a clump, the densest particle is considered to be `lead'. All other particles in the clump are considered to be `members'.  Briefly, we first sort dust particles by density.  Then, the densest particle with $\rho > \rho_\text{lead} = 2 \times 10^{-13}$~g~cm$^{-3}$ becomes the lead member of the first clump.  For successively less dense particles, we determine if their smoothing length overlaps with an existing clump and if the particle is bound to the clump.  If so and if $\rho > \rho_\text{member} = 10^{-13}$~g~cm$^{-3}$, then it is added to the clump and the clump properties are updated to account for the new member.  If not and $\rho > \rho_\text{lead}$, then it becomes the lead member of a new clump.  When a particle's smoothing length overlaps with two clumps, or if two clumps overlap, we determine if they are bound, and if so, we merge the clumps.  These values are chosen to help the clump-finding algorithm distinguish clumps from the spiral arm they reside in. The impact of our choices do not meaningfully affect the analysis of the clumps.

Since we are evaluating clump membership by inspecting particles in order of decreasing density rather than spatial proximity to a clump, we repeat this process iteratively until the number of clumps and their properties have converged.  This is required since each iteration will find more distant particles whose smoothing length will overlap with the clump.
Only dust particles have been considered for the clump finding algorithm.  Clumps with less than 58 particles (the average number of neighbours) are considered to be unresolved, and are removed.

\section{Results}
\label{sec:results}

\subsection{Dust concentration and clumping in the spiral arms}

Figure \ref{fig:50cm} shows the density evolution of gas (top) and 50cm sized dust grains (bottom) in a gravitationally unstable $0.2M_\odot$ disc at three moments in time. The left column shows the gas and dust at a time when the dust is initialised. The gas disc is stable to fragmentation. The middle and right columns show the simulation 0.44 and 0.88 outer orbits later, respectively. The 50cm dust grains initially have an average Stokes number, $\text{St} \sim 4$. In this regime, dust particles are expected to drift efficiently into pressure maxima \citep{1977Weidenschilling}. The spiral arms due to gravitational instability are regions of pressure maxima. They are also regions of gravitational potential minima which also attracts the dust. Hence, dust rapidly drifts towards and concentrates in the spiral arms in the middle and right panels of Figure \ref{fig:50cm}. The dust spirals are narrower compared to the gas spirals. Additionally, the dust is able to collapse into small clumps. 
Both these results are consistent with simulations in \cite{2004Rice,2006Rice,2012Gibbons,2014Gibbons,2016Booth,2016Shi,2021Baehr}.

Figure \ref{fig:Q_plot} shows the dust-to-gas mass ratio (top), the sound speed of the dust relative to the gas (middle), and Toomre Q of the dust disc (bottom). Given that dust is only $1\%$ of the entire disc mass, it is perhaps surprising that there is enough dust mass to become gravitationally unstable. Especially since the initial $Q_\mathrm{min}^\mathrm{dust} \sim 90$, which is far from the conditions required for gravitational instability to occur \citep[$Q<1.7$, ][]{2007Durisen}. As the disc has evolved, the local dust-to-gas mass ratio has increased from its initial value of 0.01. 
In spiral arms the dust-to-gas mass ratio has increased by an order of magnitude over the initial value. The sound speed of the dust is also lower relative to the gas by nearly an order of magnitude. Both of these factors explain why the dust disc collapses into clumps despite the initial dust profile being safely in the gravitationally stable regime. The increased surface density and lower sound speed relative to the gas both contribute to the dust disc becoming locally gravitationally unstable in the spiral arms, with $Q^\mathrm{dust}<1.7$ as shown in the bottom panel of Fig \ref{fig:Q_plot}.  

\subsection{Radial Drift}

In contrast to the gas, the dust is a pressureless fluid which results in their velocities being different.  In a smooth axisymmetric disc the drag force exerts a torque on the dust particles, resulting in dust migrating inwards towards the central star. However, in these simulations the gas structure is not axisymmetric. The spiral arms due to gravitational instabilities are local pressure maxima, and dust therefore drifts towards them. This can be seen in 
Figure \ref{fig:vr}, which shows the cross section slice of the radial velocity, $v_r$, of the dust in the $z=0$ plane. After the initial drift towards the spiral arms, the dust motion is along the spirals.

\subsection{Larger dust sizes}
\label{sec:DustSize}

Even the smallest grain size (50cm) in this study has Stokes numbers above unity. In non-self-gravitating discs in this regime, the dust is partially decoupled from the gas. However, as seen from Figure \ref{fig:50cm}, in a gravitationally unstable disc, the dust still traces the gas structure. The dust is not perfectly coupled as evident from the enhanced dust-to-gas mass ratios.

Figure \ref{fig:allStokes} shows the cross sectional slices of the dust density for all four dust sizes. Spiral structures are visible in all four panels, although to different extents. As the dust size increases, the dust spiral arms becomes more diffuse as the dust becomes less coupled. The weaker effect of the drag force with increasing dust size results in less dust density enhancement in the spirals as seen by the lack of contrast in the bottom right panel of Fig \ref{fig:allStokes}. Additionally, dust clumping is only seen for the smallest grain sizes (50 and 150cm). 
Figure \ref{fig:csStokes} shows the probability density function (PDF) of the sound speed of dust relative to the gas, $c_\text{s}^{\text{dust}}/c_\text{s}^{\text{gas}}$, for all four dust sizes. To compare $c_\text{s}^{\text{dust}}$ and $c_\text{s}^{\text{gas}}$ with each other, they were first individually interpolated onto a grid. The $c_\text{s}^{\text{dust}}$ increases as grain size increases. Both the more efficient dust trapping and decreased dust sound speed are why only the smallest grain sizes are prone to clumping.

The structure of the 150\,cm sized grains is the most similar to the gas structure. This can be observed by the peak of the PDF at $c_\text{s}^{\text{dust}} \sim c_\text{s}^{\text{gas}}$. Essentially, the dust and gas act similarly since the similar sound speeds cause the response of both fluids to any perturbations to be similar. Thus, the dust structure is closely coupled to the gas. 
At larger grain sizes, the drag force becomes less effective in damping velocity variations resulting in $c_\text{s}^{\text{dust}} > c_\text{s}^{\text{gas}}$. Hence the dust structure becomes more uncorrelated with the gas with increasing dust size. For the 5000\,cm sized grains, the spiral structures visible in the bottom right panel of Fig \ref{fig:allStokes} disappear after a few outer orbits. The PDF of the 50\,cm sized grains peaks at $c_\text{s}^{\text{dust}} < c_\text{s}^{\text{gas}}$. A broad tail extending to lower $c_\text{s}^{\text{dust}}/c_\text{s}^{\text{gas}}$ is also visible which allows dust to collapse further into clumps.  

\subsection{The role of self-gravity and drag}

It's apparent that drag plays an important role in shaping the dust as seen by the evolution of different sized dust grains. The role of gas gravity and dust self-gravity remains less clear. 
Here we focus on the 50\,cm sized grains, where the drag force is most efficient for this simulation, making it easier to study the role of self-gravity compared to the drag force. 
\begin{figure}
    \centering
    \includegraphics[width=\linewidth]{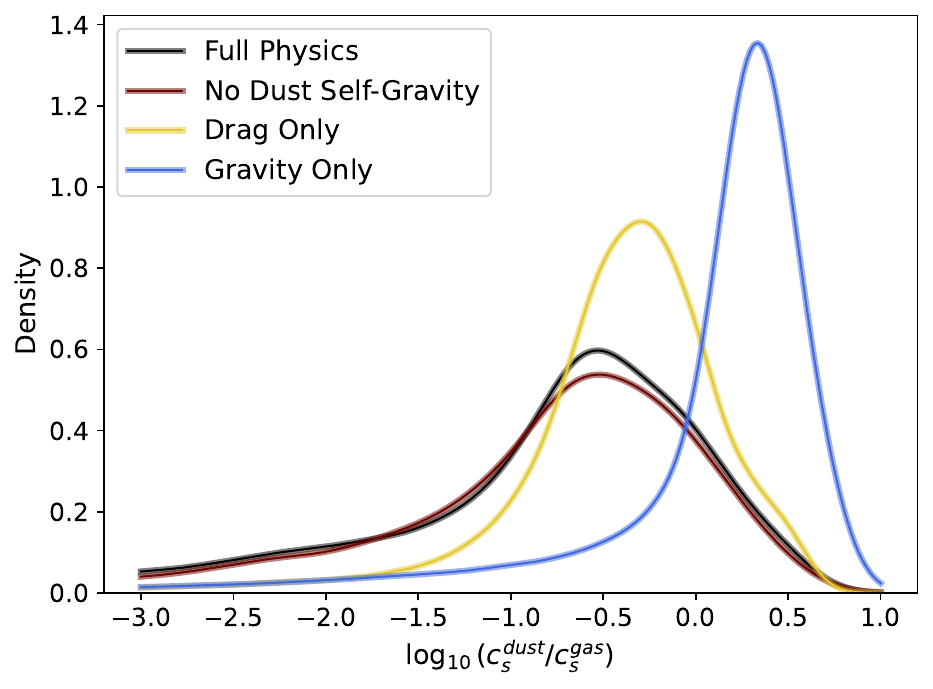}
    \caption{A probability density function (PDF) of the sound speed of dust relative to the gas, $c_\text{s}^{\text{dust}}/c_\text{s}^{\text{gas}}$, for all four simulations with varying physics as described in Figure \ref{fig:allPhysics} and in \S\ref{sec:suite}. When the dust only feels the drag force or gravity, it results in an increase in velocity dispersion resulting in dust decoupling from the gas.}
    \label{fig:csPhysics}
\end{figure} 

Figure \ref{fig:allPhysics} shows the cross sectional slices of the dust density 0.88 outer orbits after the simulation is resumed with different combinations of physics turned off.  
The reference simulation continues to evolve as expected. The dust remains trapped in the spiral arms, with numerous clumps. When the dust cannot feel the gravity of other dust particles (the ``No Dust Self-Gravity'' simulation), clumps cannot form.
The dust evolves similarly to the reference simulation otherwise with dust still being trapped by the gas spirals. 
When dust cannot feel the gas gravity either (the ``Drag Only'' simulation),
the continued presence of the drag force keeps the dust in thin filaments. However, as the dust no longer feels the gravitational forces due to the gas, it no longer traces the spiral structure (also see Figure \ref{fig:longTerm}).
Drag alone isn't sufficient to trap dust and form clumps.
Finally, when the dust cannot feel the drag force, the spiral structures visible initially form due to gravitational interactions between dust and gas. However, without drag, the spirals are not maintained and disappear after a couple of outer orbits. The dust behaves similarly to the largest dust size in \S\ref{sec:DustSize}; the dust cannot form clumps and does not trace the gas spirals. 

To illustrate the role gravity and drag have on the dust dynamics, consider the plot of the PDF of $c_\text{s}^{\text{dust}}/c_\text{s}^{\text{gas}}$ in Figure \ref{fig:csPhysics} along with the two timescales described in \S\ref{sec:DD}: the stopping time; and the orbital time. For the dust sizes (50cm, $\text{St} \approx 4$) in Figure \ref{fig:allPhysics}, the stopping time is larger than the orbital time. The spiral structures are constantly moving at the orbital velocity. The longer timescale of the drag force causes the dust to decouple when it only feels the drag force (bottom left panel in Figure \ref{fig:allPhysics}). The dust velocities are unable to decay to the gas velocities within the time they cross the gas spirals. However, the dust spirals are also regions of gravitational potential minima.    
If the dust only feels gravitational forces (bottom right panel in Figure \ref{fig:allPhysics}), the dust initially forms spiral structures but quickly decouples from the gas as there is no longer any force eliminating any differential velocity between the dust and gas resulting in $c_\text{s}^{\text{dust}} > c_\text{s}^{\text{gas}}$. This is essentially what is seen with larger dust sizes in Section \ref{sec:DD}. While neither drag nor gravity on their own is enough to trap dust, their combined influence enables the dust to keep up with the gas spirals (top right panel in Figure \ref{fig:allPhysics}).

However, this analysis is specific to the large grain sizes studied here ($\text{St} > 1$). Smaller grain sizes with $\text{St} < 0.1$ will still be able to trace the spirals due to the drag force alone: the short stopping time results in the dust velocities decaying to the gas velocities before the gas spirals have moved significantly \citep{2021Baehr}.  

As mentioned earlier, all dust sizes here are in the regime where dust is expected to be decoupled from the gas ($\text{St} > 1$). So it's perhaps a bit surprising that the dust was still able to trace the gas spirals. The gravity of the gas is usually negligible in most studies of dusty protoplanetary discs which are gravitationally stable. These results show that in a gravitationally unstable disc, the combined effect of gas gravity and drag is necessary for large dust grains to be well coupled to the gas and trace the spirals. For the dust to form clumps, dust self-gravity is necessary. Although including full physics results in a simulation that is at least an order of magnitude slower to compute (without drag), it is important to do so, particularly for studying the formation of dust clumps.

\subsection{Clump formation}

\begin{figure}
    \centering
    \includegraphics[width=\linewidth]{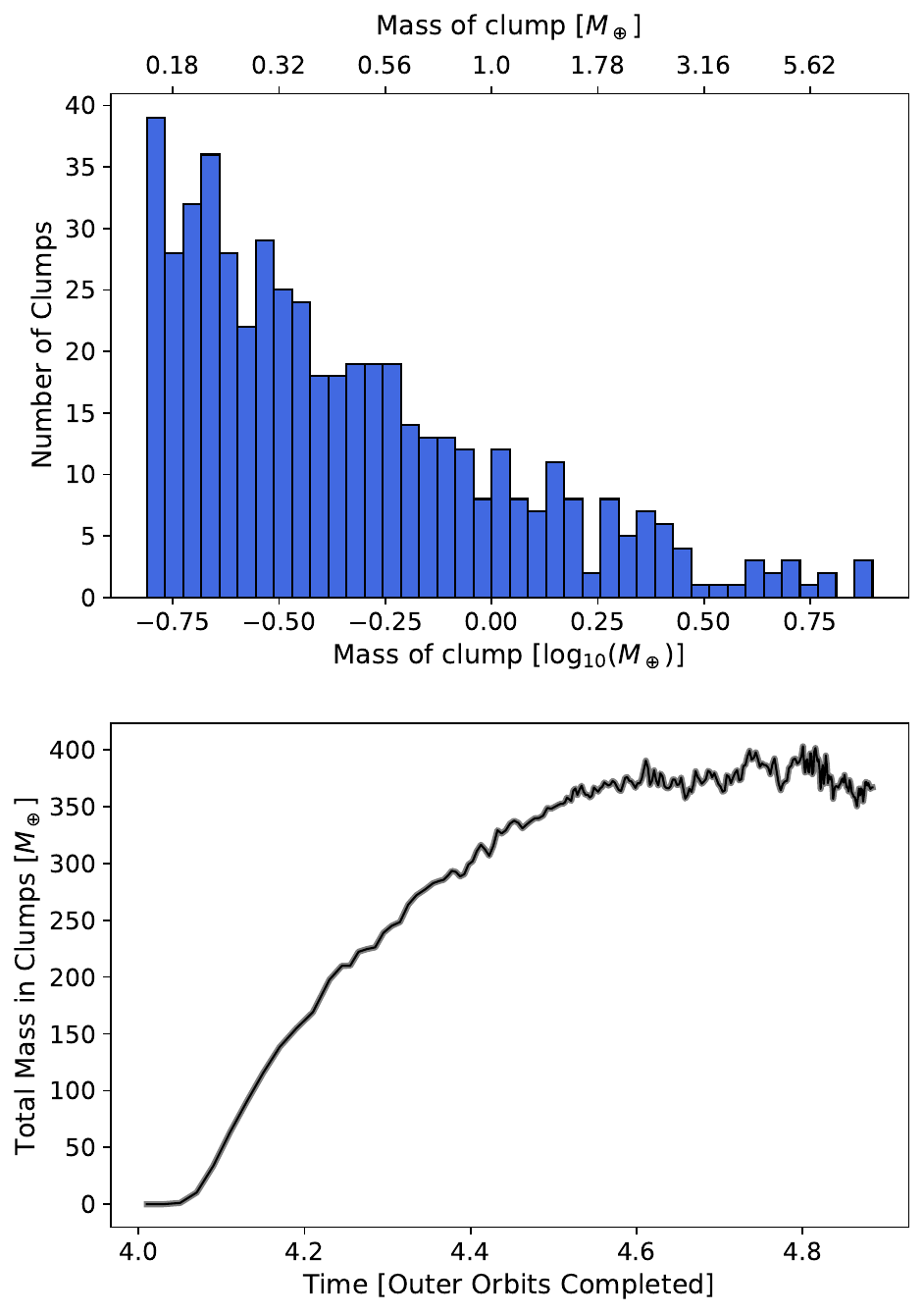}
    \caption{The top panel shows a histogram of the clump mass at $t=4.88$ outer orbits. A majority of the clumps are less than a few Earth masses, and hence might not be massive enough to become giant planets through runaway gas accretion. The bottom panel shows the time evolution of the total mass found in clumps. Within half an outer orbit $60\%$ of the dust disc ($400M_\oplus$) is found in clumps as shown by the black line (Full Physics simulation). Only resolved clumps are plotted.}
    \label{fig:clumpAnalysis}
\end{figure}

The top panel of Figure \ref{fig:clumpAnalysis} shows a histogram of all resolved clump masses at $t=4.88$ outer orbits for the reference (Full Physics) simulation where the dust feels the gravity of both the gas and dust, and the drag force. Figure \ref{fig:clumps} shows the locations of all resolved clumps. The clumps range from $0.15M_\oplus$ (the resolution limit) to ${\sim}6 M_\oplus$, with a high majority of the clumps being less than the critical mass (${\sim}5M_\oplus$) to undergo runaway gas accretion \citep{1999Papaloizou,2003Rice}. The critical mass is dependent on the disc parameters, and solid accretion rate and composition \citep[see review by][]{2022Drazkowska}.
The lower panel of Figure \ref{fig:clumpAnalysis} shows the total mass found in gravitationally bound clumps as a function of time  Clump formation is rapid with roughly a Jupiter mass of dust accumulating in clumps within half an outer orbit. The total mass in clumps steadies after this initial rapid build-up of clumps once roughly half the dust is in clumps.

\section{Discussion}
\label{sec:Discussion}

\subsection{Limitations}

The main limitations of this study involve how the dust is modelled. The dust used in the simulations in this study is a single fixed size. While this simplifies the simulations, it is not representative of reality where dust is made up of a mixture of different sizes and can grow. Assuming all the dust is of the same size (and therefore all the dust mass is in a single grain size, rather than a power-law distribution) is the optimal scenario for the dust disc to become gravitationally unstable and form clumps. 
Previous studies have shown the dust only begins to form clumps when St $>0.3$ \citep{2016Booth,2012Gibbons}. Thus if the mass of dust was spread across various dust sizes, then there might not be enough mass for dust of a given size to become gravitationally unstable. With a range of dust sizes, and hence Stokes numbers, the effectiveness of dust trapping will be varied. Not all dust will be optimally trapped, thus resulting in reduced dust density enhancements and increasing $Q$.

However, including dust growth could aid dust clumping. Small dust grains are well coupled to the gas and collect in the gas spirals \citep{2021Baehr,2012Gibbons}. In these regions, \cite{2020Elbakyan} showed that dust can rapidly grow from micron sized grains to a few centimetres with St $\sim 0.3$. If enough dust can grow to this size where they are most strongly influenced by the gas spirals, then the dust might still be able to become gravitationally unstable and form clumps.

Our simulations have not included dust back-reaction onto the gas. With the enhancement of the dust-to-gas mass ratios as seen in Figure \ref{fig:Q_plot}, this could become relevant. However, based on shearing box simulations in \cite{2014Gibbons} and \cite{2022Baehr}, the qualitative nature of the role of drag and gravitational forces as described in this study is likely to be unchanged if back-reaction was included. However, the mass of the clumps measured would be impacted. \cite{2022Baehr} found clump masses to be smaller, but more numerous when back-reaction was included. 

These simulations also made use of a simplifying approximation to the stopping time. This simplification allowed for the drag force term in Eq \ref{eq:DD} to be independent of the dust density which enables the simulations to evolve for a longer time after dust clumps are formed. Our results of clump formation still compare well with \cite{2006Rice} which is similarly as rapid. While their simulations stopped when 15\% of dust was in clumps, the amount of dust in clumps was still rising. As we could evolve further, we find that the dust continued to accumulate in clumps until ${\sim}50\%$ of the dust was in clumps. We find the clump masses (${\sim}0.1-5M_\oplus$) are slightly higher than the results in \cite{2022Baehr}, but similar to \cite{2023Longarini,2023bLongarini}, although this could be due to the lower Stokes numbers in the former's simulations. Additionally they utilised a fixed stopping time, whereas we used a fixed particle size which results in the stopping time varying with the particle location. A fixed particle size has been shown to result in stronger dust concentrations \citep{2016Shi}.

\subsection{Planet formation}
The simulations here show that planet formation in gravitationally unstable discs is not restricted to gas giants. The spiral arms of gravitationally unstable discs provide conditions that favour dust concentration which rapidly collapses to form numerous Earth mass clumps. The short timescale at which the Earth mass clumps form could help explain how some of these discs exhibit ring \& gap structure (often associated with evidence of planet formation) despite being less than a million years old \citep{2015ALMA,2018Sheehan,Segura-Cox2020}. The clumps formed in gravitationally unstable discs could become the seeds of the planets that go on to carve rings \& gaps in ALMA observations. This work adds to the idea that gravitationally unstable discs may form planets lower in mass than giant planets \citep{2021Deng}.

The simulations in this study highlight that dust trapping in the spiral arms of \textit{non-fragmenting} gravitationally unstable discs is a viable mechanism for forming planetesimals and/or planetary cores. Most discs are expected to go through a gravitationally unstable phase during their evolution. Thus, planet formation in very young discs could be explained by a population of planetesimals formed in the earliest stages of a disc's evolution \citep{2018Nixon}. However, the runtime of these simulations is restricted as the minimum timestep required for accurate evolution decreases as dust particles get closer together as it clumps. Future global simulations with dust growth are necessary to evolve these simulations further to investigate whether the clumps formed here remain as Earth to Super-Earths or whether they eventually become giant planets. Recently, \cite{2023Baehr} have explored this question, though not with global simulations.

\section{conclusion}

We perform 3D SPH simulations to investigate the role of gravitational and drag forces on dust concentration in gravitationally unstable protoplanetary discs. In the regime studied in this work (large dust grains with Stokes numbers greater than unity), our simulations show that both drag and gravity are necessary for dust to be well coupled to the gas and trace the spirals. Neither on their own is capable.

If the dust is strongly influenced by the drag force, the density increases while the sound speed of the dust decreases. The combined effect results in the dust becoming gravitationally unstable and forming gravitationally bound clumps. The speed at which the dust forms Earth to Super-Earth mass clumps is rapid. Within half an outer orbit, roughly half the dust disc is in the form of clumps. After which, clump formation ceases (as the disc runs out of material to form more). While dust is still trapped in the gas spirals without dust self-gravity, it is unable to collapse to form clumps. Neglecting either gas gravity or drag results in the dust being unable to trace the gas spirals.

The results presented here show that planet formation through gravitational instability may not be limited to just giant planets. If the clumps formed remain under the critical mass for runaway gas accretion (${\sim}5M_\oplus$), then gravitational instability in discs stable to gas fragmentation could form Earth to Super-Earths very quickly through efficient dust trapping in the gas spirals.

\section*{Data Availability}

The data from the simulations used to create all plots in this article is available on reasonable request to the corresponding author. The software used to create and visualise the simulations are publicly available:
\begin{description}[leftmargin=5em,style=nextline,font=\normalfont\normalsize]
    \item[\textsc{Phantom}] \url{https://github.com/danieljprice/phantom} \citep{2018Price}
    \item[\textsc{Splash}] \url{https://github.com/danieljprice/splash} \citep{2007Price}
    \item[\textsc{Sarracen}] \url{https://github.com/ttricco/sarracen}\\ \citep{sarracen}
\end{description}

\section*{Acknowledgements}

S.R. acknowledges support from a Royal Society Enhancement Award, and (along with R.A.) funding from the Science \& Technology Facilities Council (STFC) through Consolidated Grant ST/W000857/1. F.M. acknowledges support from the Royal Society Dorothy Hodgkin Fellowship. R.N. acknowledges support from UKRI/EPSRC through a Stephen Hawking Fellowship (EP/T017287/1). J.W. acknowledges support from the University of St Andrews. H.A. acknowledges funding from the European Research Council (ERC) under the European Union’s Horizon 2020 research and innovation programme (grant agreement No 101054502). K.R. is grateful for support from the UK STFC via grant ST/V000594/1. R.A.B is supported by a Royal Society University Research Fellowship. S.R. would like to thank Cristiano Longarini and Grant Kennedy for useful discussions, and the anonymous referee for their useful comments which benefited this work. The simulations were performed using Orac \& Avon, the High Performance Computing clusters at the University of Warwick, and ALICE, the High Performance Computing Facility at the University of Leicester.



\bibliographystyle{mnras}
\bibliography{DustGI_Disc} 

\begin{thebibliography}{}
\makeatletter
\relax
\def\mn@urlcharsother{\let\do\@makeother \do\$\do\&\do\#\do\^\do\_\do\%\do\~}
\def\mn@doi{\begingroup\mn@urlcharsother \@ifnextchar [ {\mn@doi@}
  {\mn@doi@[]}}
\def\mn@doi@[#1]#2{\def\@tempa{#1}\ifx\@tempa\@empty \href
  {http://dx.doi.org/#2} {doi:#2}\else \href {http://dx.doi.org/#2}
  {\color{violet}#1}\fi \endgroup}
\def\mn@eprint#1#2{\mn@eprint@#1:#2::\@nil}
\def\mn@eprint@arXiv#1{\href {http://arxiv.org/abs/#1} {{\tt arXiv:#1}}}
\def\mn@eprint@dblp#1{\href {http://dblp.uni-trier.de/rec/bibtex/#1.xml}
  {dblp:#1}}
\def\mn@eprint@#1:#2:#3:#4\@nil{\def\@tempa {#1}\def\@tempb {#2}\def\@tempc
  {#3}\ifx \@tempc \@empty \let \@tempc \@tempb \let \@tempb \@tempa \fi \ifx
  \@tempb \@empty \def\@tempb {arXiv}\fi \@ifundefined
  {mn@eprint@\@tempb}{\@tempb:\@tempc}{\expandafter \expandafter \csname
  mn@eprint@\@tempb\endcsname \expandafter{\@tempc}}}

\bibitem[\protect\citeauthoryear{{ALMA Partnership} et~al.,}{{ALMA Partnership}
  et~al.}{2015}]{2015ALMA}
{ALMA Partnership} et~al., 2015, \mn@doi [\apj] {10.1088/2041-8205/808/1/L3},
  \href {https://ui.adsabs.harvard.edu/abs/2015ApJ...808L...3A} {808, L3}

\bibitem[\protect\citeauthoryear{{Andrews} et~al.,}{{Andrews}
  et~al.}{2016}]{2016Andrews}
{Andrews} S.~M.,  et~al., 2016, \mn@doi [\apjl] {10.3847/2041-8205/820/2/L40},
  \href {https://ui.adsabs.harvard.edu/abs/2016ApJ...820L..40A} {820, L40}

\bibitem[\protect\citeauthoryear{{Andrews} et~al.,}{{Andrews}
  et~al.}{2018}]{2018Andrews}
{Andrews} S.~M.,  et~al., 2018, \mn@doi [\apj] {10.3847/2041-8213/aaf741},
  \href {https://ui.adsabs.harvard.edu/abs/2018ApJ...869L..41A} {869, L41}

\bibitem[\protect\citeauthoryear{{Baehr}}{{Baehr}}{2023}]{2023Baehr}
{Baehr} H.,  2023, \mn@doi [\mnras] {10.1093/mnras/stad1564}, \href
  {https://ui.adsabs.harvard.edu/abs/2023MNRAS.523.3348B} {523, 3348}

\bibitem[\protect\citeauthoryear{{Baehr} \& {Zhu}}{{Baehr} \&
  {Zhu}}{2021}]{2021Baehr}
{Baehr} H.,  {Zhu} Z.,  2021, \mn@doi [\apj] {10.3847/1538-4357/abddb3}, \href
  {https://ui.adsabs.harvard.edu/abs/2021ApJ...909..135B} {909, 135}

\bibitem[\protect\citeauthoryear{{Baehr}, {Zhu}  \& {Yang}}{{Baehr}
  et~al.}{2022}]{2022Baehr}
{Baehr} H.,  {Zhu} Z.,   {Yang} C.-C.,  2022, \mn@doi [\apj]
  {10.3847/1538-4357/ac7228}, \href
  {https://ui.adsabs.harvard.edu/abs/2022ApJ...933..100B} {933, 100}

\bibitem[\protect\citeauthoryear{{Bate}, {Bonnell}  \& {Price}}{{Bate}
  et~al.}{1995}]{1995Bate}
{Bate} M.~R.,  {Bonnell} I.~A.,   {Price} N.~M.,  1995, \mn@doi [\mnras]
  {10.1093/mnras/277.2.362}, \href
  {https://ui.adsabs.harvard.edu/abs/1995MNRAS.277..362B} {277, 362}

\bibitem[\protect\citeauthoryear{{Booth} \& {Clarke}}{{Booth} \&
  {Clarke}}{2016}]{2016Booth}
{Booth} R.~A.,  {Clarke} C.~J.,  2016, \mn@doi [\mnras] {10.1093/mnras/stw488},
  \href {https://ui.adsabs.harvard.edu/abs/2016MNRAS.458.2676B} {458, 2676}

\bibitem[\protect\citeauthoryear{{Booth} \& {Ilee}}{{Booth} \&
  {Ilee}}{2020}]{2020Booth}
{Booth} A.~S.,  {Ilee} J.~D.,  2020, \mn@doi [\mnras] {10.1093/mnrasl/slaa014},
  \href {https://ui.adsabs.harvard.edu/abs/2020MNRAS.493L.108B} {493, L108}

\bibitem[\protect\citeauthoryear{{Boss}}{{Boss}}{1997}]{1997Boss}
{Boss} A.~P.,  1997, \mn@doi [Science] {10.1126/science.276.5320.1836}, \href
  {https://ui.adsabs.harvard.edu/abs/1997Sci...276.1836B} {276, 1836}

\bibitem[\protect\citeauthoryear{{Cadman}, {Hall}, {Rice}, {Harries}  \&
  {Klaassen}}{{Cadman} et~al.}{2020}]{2020Cadman}
{Cadman} J.,  {Hall} C.,  {Rice} K.,  {Harries} T.~J.,   {Klaassen} P.~D.,
  2020, \mn@doi [\mnras] {10.1093/mnras/staa2596}, \href
  {https://ui.adsabs.harvard.edu/abs/2020MNRAS.498.4256C} {498, 4256}

\bibitem[\protect\citeauthoryear{{Calcino}, {Hilder}, {Price}, {Pinte},
  {Bollati}, {Lodato}  \& {Norfolk}}{{Calcino} et~al.}{2022}]{2021Calcino}
{Calcino} J.,  {Hilder} T.,  {Price} D.~J.,  {Pinte} C.,  {Bollati} F.,
  {Lodato} G.,   {Norfolk} B.~J.,  2022, \mn@doi [\apjl]
  {10.3847/2041-8213/ac64a7}, \href
  {https://ui.adsabs.harvard.edu/abs/2022ApJ...929L..25C} {929, L25}

\bibitem[\protect\citeauthoryear{{Cullen} \& {Dehnen}}{{Cullen} \&
  {Dehnen}}{2010}]{2010Cullen}
{Cullen} L.,  {Dehnen} W.,  2010, \mn@doi [\mnras]
  {10.1111/j.1365-2966.2010.17158.x}, \href
  {https://ui.adsabs.harvard.edu/abs/2010MNRAS.408..669C} {408, 669}

\bibitem[\protect\citeauthoryear{{Deng}, {Mayer}  \& {Helled}}{{Deng}
  et~al.}{2021}]{2021Deng}
{Deng} H.,  {Mayer} L.,   {Helled} R.,  2021, \mn@doi [Nature Astronomy]
  {10.1038/s41550-020-01297-6}, \href
  {https://ui.adsabs.harvard.edu/abs/2021NatAs...5..440D} {5, 440}

\bibitem[\protect\citeauthoryear{{Dipierro}, {Price}, {Laibe}, {Hirsh},
  {Cerioli}  \& {Lodato}}{{Dipierro} et~al.}{2015}]{2015Dipierro}
{Dipierro} G.,  {Price} D.,  {Laibe} G.,  {Hirsh} K.,  {Cerioli} A.,   {Lodato}
  G.,  2015, \mn@doi [\mnras] {10.1093/mnrasl/slv105}, \href
  {https://ui.adsabs.harvard.edu/abs/2015MNRAS.453L..73D} {453, L73}

\bibitem[\protect\citeauthoryear{{Dipierro} et~al.,}{{Dipierro}
  et~al.}{2018}]{2018Dipierro}
{Dipierro} G.,  et~al., 2018, \mn@doi [\mnras] {10.1093/mnras/sty181}, \href
  {https://ui.adsabs.harvard.edu/abs/2018MNRAS.475.5296D} {475, 5296}

\bibitem[\protect\citeauthoryear{{Dr{\k{a}}{\.z}kowska}
  et~al.,}{{Dr{\k{a}}{\.z}kowska} et~al.}{2023}]{2022Drazkowska}
{Dr{\k{a}}{\.z}kowska} J.,  et~al., 2023, \mn@doi [Astronomical Society of the
  Pacific Conference Series] {10.48550/arXiv.2203.09759}, \href
  {https://ui.adsabs.harvard.edu/abs/2023ASPC..534..717D} {534, 717}

\bibitem[\protect\citeauthoryear{{Durisen}, {Boss}, {Mayer}, {Nelson}, {Quinn}
  \& {Rice}}{{Durisen} et~al.}{2007}]{2007Durisen}
{Durisen} R.~H.,  {Boss} A.~P.,  {Mayer} L.,  {Nelson} A.~F.,  {Quinn} T.,
  {Rice} W.~K.~M.,  2007, Protostars and Planets V, \href
  {https://ui.adsabs.harvard.edu/abs/2007prpl.conf..607D} {p.~607}

\bibitem[\protect\citeauthoryear{{Elbakyan}, {Johansen}, {Lambrechts},
  {Akimkin}  \& {Vorobyov}}{{Elbakyan} et~al.}{2020}]{2020Elbakyan}
{Elbakyan} V.~G.,  {Johansen} A.,  {Lambrechts} M.,  {Akimkin} V.,   {Vorobyov}
  E.~I.,  2020, \mn@doi [\aap] {10.1051/0004-6361/201937198}, \href
  {https://ui.adsabs.harvard.edu/abs/2020A&A...637A...5E} {637, A5}

\bibitem[\protect\citeauthoryear{{Fedele} et~al.,}{{Fedele}
  et~al.}{2018}]{2018Fedele}
{Fedele} D.,  et~al., 2018, \mn@doi [\aap] {10.1051/0004-6361/201731978}, \href
  {https://ui.adsabs.harvard.edu/abs/2018A&A...610A..24F} {610, A24}

\bibitem[\protect\citeauthoryear{{Gammie}}{{Gammie}}{2001}]{2001Gammie}
{Gammie} C.~F.,  2001, \mn@doi [\apj] {10.1086/320631}, \href
  {https://ui.adsabs.harvard.edu/abs/2001ApJ...553..174G} {553, 174}

\bibitem[\protect\citeauthoryear{{Gibbons}, {Rice}  \&
  {Mamatsashvili}}{{Gibbons} et~al.}{2012}]{2012Gibbons}
{Gibbons} P.~G.,  {Rice} W.~K.~M.,   {Mamatsashvili} G.~R.,  2012, \mn@doi
  [\mnras] {10.1111/j.1365-2966.2012.21731.x}, \href
  {https://ui.adsabs.harvard.edu/abs/2012MNRAS.426.1444G} {426, 1444}

\bibitem[\protect\citeauthoryear{{Gibbons}, {Mamatsashvili}  \&
  {Rice}}{{Gibbons} et~al.}{2014}]{2014Gibbons}
{Gibbons} P.~G.,  {Mamatsashvili} G.~R.,   {Rice} W.~K.~M.,  2014, \mn@doi
  [\mnras] {10.1093/mnras/stu809}, \href
  {https://ui.adsabs.harvard.edu/abs/2014MNRAS.442..361G} {442, 361}

\bibitem[\protect\citeauthoryear{{Hall} et~al.,}{{Hall}
  et~al.}{2020}]{2020Hall}
{Hall} C.,  et~al., 2020, \mn@doi [\apj] {10.3847/1538-4357/abac17}, \href
  {https://ui.adsabs.harvard.edu/abs/2020ApJ...904..148H} {904, 148}

\bibitem[\protect\citeauthoryear{Harris \& Tricco}{Harris \&
  Tricco}{2023}]{sarracen}
Harris A.,  Tricco T.~S.,  2023, \mn@doi [Journal of Open Source Software]
  {10.21105/joss.05263}, 8, 5263

\bibitem[\protect\citeauthoryear{{Huang} et~al.,}{{Huang}
  et~al.}{2018a}]{2018Huang}
{Huang} J.,  et~al., 2018a, \mn@doi [\apjl] {10.3847/2041-8213/aaf740}, \href
  {https://ui.adsabs.harvard.edu/abs/2018ApJ...869L..42H} {869, L42}

\bibitem[\protect\citeauthoryear{{Huang} et~al.,}{{Huang}
  et~al.}{2018b}]{2018bHuang}
{Huang} J.,  et~al., 2018b, \mn@doi [\apjl] {10.3847/2041-8213/aaf7a0}, \href
  {https://ui.adsabs.harvard.edu/abs/2018ApJ...869L..43H} {869, L43}

\bibitem[\protect\citeauthoryear{{Kley} \& {Nelson}}{{Kley} \&
  {Nelson}}{2012}]{2012Kley}
{Kley} W.,  {Nelson} R.~P.,  2012, \mn@doi [\araa]
  {10.1146/annurev-astro-081811-125523}, \href
  {https://ui.adsabs.harvard.edu/abs/2012ARA&A..50..211K} {50, 211}

\bibitem[\protect\citeauthoryear{{Kwok}}{{Kwok}}{1975}]{1975Kwok}
{Kwok} S.,  1975, \mn@doi [\apj] {10.1086/153637}, \href
  {https://ui.adsabs.harvard.edu/abs/1975ApJ...198..583K} {198, 583}

\bibitem[\protect\citeauthoryear{{Laibe} \& {Price}}{{Laibe} \&
  {Price}}{2012a}]{2012aLaibe}
{Laibe} G.,  {Price} D.~J.,  2012a, \mn@doi [\mnras]
  {10.1111/j.1365-2966.2011.20202.x}, \href
  {https://ui.adsabs.harvard.edu/abs/2012MNRAS.420.2345L} {420, 2345}

\bibitem[\protect\citeauthoryear{{Laibe} \& {Price}}{{Laibe} \&
  {Price}}{2012b}]{2012bLaibe}
{Laibe} G.,  {Price} D.~J.,  2012b, \mn@doi [\mnras]
  {10.1111/j.1365-2966.2011.20201.x}, \href
  {https://ui.adsabs.harvard.edu/abs/2012MNRAS.420.2365L} {420, 2365}

\bibitem[\protect\citeauthoryear{{Lodato} et~al.,}{{Lodato}
  et~al.}{2019}]{2019Lodato}
{Lodato} G.,  et~al., 2019, \mn@doi [\mnras] {10.1093/mnras/stz913}, \href
  {https://ui.adsabs.harvard.edu/abs/2019MNRAS.486..453L} {486, 453}

\bibitem[\protect\citeauthoryear{{Long} et~al.,}{{Long}
  et~al.}{2018}]{2018Long}
{Long} F.,  et~al., 2018, \mn@doi [\apj] {10.3847/1538-4357/aae8e1}, \href
  {https://ui.adsabs.harvard.edu/abs/2018ApJ...869...17L} {869, 17}

\bibitem[\protect\citeauthoryear{{Longarini}, {Lodato}, {Toci}, {Veronesi},
  {Hall}, {Dong}  \& {Patrick Terry}}{{Longarini} et~al.}{2021}]{2021Longarini}
{Longarini} C.,  {Lodato} G.,  {Toci} C.,  {Veronesi} B.,  {Hall} C.,  {Dong}
  R.,   {Patrick Terry} J.,  2021, \mn@doi [\apjl] {10.3847/2041-8213/ac2df6},
  \href {https://ui.adsabs.harvard.edu/abs/2021ApJ...920L..41L} {920, L41}

\bibitem[\protect\citeauthoryear{{Longarini}, {Lodato}, {Bertin}  \&
  {Armitage}}{{Longarini} et~al.}{2023a}]{2023Longarini}
{Longarini} C.,  {Lodato} G.,  {Bertin} G.,   {Armitage} P.~J.,  2023a, \mn@doi
  [\mnras] {10.1093/mnras/stac3653}, \href
  {https://ui.adsabs.harvard.edu/abs/2023MNRAS.519.2017L} {519, 2017}

\bibitem[\protect\citeauthoryear{{Longarini}, {Armitage}, {Lodato}, {Price}  \&
  {Ceppi}}{{Longarini} et~al.}{2023b}]{2023bLongarini}
{Longarini} C.,  {Armitage} P.~J.,  {Lodato} G.,  {Price} D.~J.,   {Ceppi} S.,
  2023b, \mn@doi [\mnras] {10.1093/mnras/stad1400}, \href
  {https://ui.adsabs.harvard.edu/abs/2023MNRAS.522.6217L} {522, 6217}

\bibitem[\protect\citeauthoryear{{Meru}, {Juh{\'a}sz}, {Ilee}, {Clarke},
  {Rosotti}  \& {Booth}}{{Meru} et~al.}{2017}]{2017Meru}
{Meru} F.,  {Juh{\'a}sz} A.,  {Ilee} J.~D.,  {Clarke} C.~J.,  {Rosotti} G.~P.,
   {Booth} R.~A.,  2017, \mn@doi [\apjl] {10.3847/2041-8213/aa6837}, \href
  {https://ui.adsabs.harvard.edu/abs/2017ApJ...839L..24M} {839, L24}

\bibitem[\protect\citeauthoryear{{Nealon}, {Price}  \& {Nixon}}{{Nealon}
  et~al.}{2015}]{2015Nealon}
{Nealon} R.,  {Price} D.~J.,   {Nixon} C.~J.,  2015, \mn@doi [\mnras]
  {10.1093/mnras/stv014}, \href
  {https://ui.adsabs.harvard.edu/abs/2015MNRAS.448.1526N} {448, 1526}

\bibitem[\protect\citeauthoryear{{Nixon}, {King}  \& {Pringle}}{{Nixon}
  et~al.}{2018}]{2018Nixon}
{Nixon} C.~J.,  {King} A.~R.,   {Pringle} J.~E.,  2018, \mn@doi [\mnras]
  {10.1093/mnras/sty593}, \href
  {https://ui.adsabs.harvard.edu/abs/2018MNRAS.477.3273N} {477, 3273}

\bibitem[\protect\citeauthoryear{{Paardekooper}, {Dong}, {Duffell}, {Fung},
  {Masset}, {Ogilvie}  \& {Tanaka}}{{Paardekooper} et~al.}{2022}]{2022Sijme}
{Paardekooper} S.-J.,  {Dong} R.,  {Duffell} P.,  {Fung} J.,  {Masset} F.~S.,
  {Ogilvie} G.,   {Tanaka} H.,  2022, arXiv e-prints, \href
  {https://ui.adsabs.harvard.edu/abs/2022arXiv220309595P} {p. arXiv:2203.09595}

\bibitem[\protect\citeauthoryear{{Paneque-Carre{\~n}o}
  et~al.,}{{Paneque-Carre{\~n}o} et~al.}{2021}]{2021Paneque}
{Paneque-Carre{\~n}o} T.,  et~al., 2021, \mn@doi [\apj]
  {10.3847/1538-4357/abf243}, \href
  {https://ui.adsabs.harvard.edu/abs/2021ApJ...914...88P} {914, 88}

\bibitem[\protect\citeauthoryear{{Papaloizou} \& {Terquem}}{{Papaloizou} \&
  {Terquem}}{1999}]{1999Papaloizou}
{Papaloizou} J. C.~B.,  {Terquem} C.,  1999, \mn@doi [\apj] {10.1086/307581},
  \href {https://ui.adsabs.harvard.edu/abs/1999ApJ...521..823P} {521, 823}

\bibitem[\protect\citeauthoryear{{Perez}, {Dunhill}, {Casassus}, {Roman},
  {Szul{\'a}gyi}, {Flores}, {Marino}  \& {Montesinos}}{{Perez}
  et~al.}{2015}]{2015Perez}
{Perez} S.,  {Dunhill} A.,  {Casassus} S.,  {Roman} P.,  {Szul{\'a}gyi} J.,
  {Flores} C.,  {Marino} S.,   {Montesinos} M.,  2015, \mn@doi [\apjl]
  {10.1088/2041-8205/811/1/L5}, \href
  {https://ui.adsabs.harvard.edu/abs/2015ApJ...811L...5P} {811, L5}

\bibitem[\protect\citeauthoryear{{P{\'e}rez} et~al.,}{{P{\'e}rez}
  et~al.}{2016}]{2016Perez}
{P{\'e}rez} L.~M.,  et~al., 2016, \mn@doi [Science] {10.1126/science.aaf8296},
  \href {https://ui.adsabs.harvard.edu/abs/2016Sci...353.1519P} {353, 1519}

\bibitem[\protect\citeauthoryear{{Pinte} et~al.,}{{Pinte}
  et~al.}{2018}]{2018bPinte}
{Pinte} C.,  et~al., 2018, \mn@doi [\apjl] {10.3847/2041-8213/aac6dc}, \href
  {https://ui.adsabs.harvard.edu/abs/2018ApJ...860L..13P} {860, L13}

\bibitem[\protect\citeauthoryear{{Pinte} et~al.,}{{Pinte}
  et~al.}{2019}]{2019Pinte}
{Pinte} C.,  et~al., 2019, \mn@doi [Nature Astronomy]
  {10.1038/s41550-019-0852-6}, \href
  {https://ui.adsabs.harvard.edu/abs/2019NatAs...3.1109P} {3, 1109}

\bibitem[\protect\citeauthoryear{{Pinte} et~al.,}{{Pinte}
  et~al.}{2020}]{2020Pinte}
{Pinte} C.,  et~al., 2020, \mn@doi [\apjl] {10.3847/2041-8213/ab6dda}, \href
  {https://ui.adsabs.harvard.edu/abs/2020ApJ...890L...9P} {890, L9}

\bibitem[\protect\citeauthoryear{{Pinte} et~al.,}{{Pinte}
  et~al.}{2023a}]{2023Pinte}
{Pinte} C.,  et~al., 2023a, \mn@doi [\mnras] {10.1093/mnrasl/slad010}, \href
  {https://ui.adsabs.harvard.edu/abs/2023MNRAS.526L..41P} {526, L41}

\bibitem[\protect\citeauthoryear{{Pinte}, {Teague}, {Flaherty}, {Hall},
  {Facchini}  \& {Casassus}}{{Pinte} et~al.}{2023b}]{2022Pinte}
{Pinte} C.,  {Teague} R.,  {Flaherty} K.,  {Hall} C.,  {Facchini} S.,
  {Casassus} S.,  2023b, \mn@doi [Astronomical Society of the Pacific
  Conference Series] {10.48550/arXiv.2203.09528}, \href
  {https://ui.adsabs.harvard.edu/abs/2023ASPC..534..645P} {534, 645}

\bibitem[\protect\citeauthoryear{{Price}}{{Price}}{2007}]{2007Price}
{Price} D.~J.,  2007, \mn@doi [\pasa] {10.1071/AS07022}, \href
  {https://ui.adsabs.harvard.edu/abs/2007PASA...24..159P} {24, 159}

\bibitem[\protect\citeauthoryear{{Price} \& {Laibe}}{{Price} \&
  {Laibe}}{2020}]{2020Price}
{Price} D.~J.,  {Laibe} G.,  2020, \mn@doi [\mnras] {10.1093/mnras/staa1366},
  \href {https://ui.adsabs.harvard.edu/abs/2020MNRAS.495.3929P} {495, 3929}

\bibitem[\protect\citeauthoryear{{Price} et~al.,}{{Price}
  et~al.}{2018}]{2018Price}
{Price} D.~J.,  et~al., 2018, \mn@doi [\pasa] {10.1017/pasa.2018.25}, \href
  {https://ui.adsabs.harvard.edu/abs/2018PASA...35...31P} {35, e031}

\bibitem[\protect\citeauthoryear{{Rice} \& {Armitage}}{{Rice} \&
  {Armitage}}{2003}]{2003Rice}
{Rice} W.~K.~M.,  {Armitage} P.~J.,  2003, \mn@doi [\apjl] {10.1086/380390},
  \href {https://ui.adsabs.harvard.edu/abs/2003ApJ...598L..55R} {598, L55}

\bibitem[\protect\citeauthoryear{{Rice}, {Lodato}, {Pringle}, {Armitage}  \&
  {Bonnell}}{{Rice} et~al.}{2004}]{2004Rice}
{Rice} W.~K.~M.,  {Lodato} G.,  {Pringle} J.~E.,  {Armitage} P.~J.,   {Bonnell}
  I.~A.,  2004, \mn@doi [\mnras] {10.1111/j.1365-2966.2004.08339.x}, \href
  {https://ui.adsabs.harvard.edu/abs/2004MNRAS.355..543R} {355, 543}

\bibitem[\protect\citeauthoryear{{Rice}, {Lodato}, {Pringle}, {Armitage}  \&
  {Bonnell}}{{Rice} et~al.}{2006}]{2006Rice}
{Rice} W.~K.~M.,  {Lodato} G.,  {Pringle} J.~E.,  {Armitage} P.~J.,   {Bonnell}
  I.~A.,  2006, \mn@doi [\mnras] {10.1111/j.1745-3933.2006.00215.x}, \href
  {https://ui.adsabs.harvard.edu/abs/2006MNRAS.372L...9R} {372, L9}

\bibitem[\protect\citeauthoryear{{Riols}, {Roux}, {Latter}  \& {Lesur}}{{Riols}
  et~al.}{2020}]{2020Riols}
{Riols} A.,  {Roux} B.,  {Latter} H.,   {Lesur} G.,  2020, \mn@doi [\mnras]
  {10.1093/mnras/staa567}, \href
  {https://ui.adsabs.harvard.edu/abs/2020MNRAS.493.4631R} {493, 4631}

\bibitem[\protect\citeauthoryear{{Rowther} \& {Meru}}{{Rowther} \&
  {Meru}}{2020}]{2020Rowther}
{Rowther} S.,  {Meru} F.,  2020, \mn@doi [\mnras] {10.1093/mnras/staa1590},
  \href {https://ui.adsabs.harvard.edu/abs/2020MNRAS.496.1598R} {496, 1598}

\bibitem[\protect\citeauthoryear{{Rowther}, {Meru}, {Kennedy}, {Nealon}  \&
  {Pinte}}{{Rowther} et~al.}{2020}]{2020bRowther}
{Rowther} S.,  {Meru} F.,  {Kennedy} G.~M.,  {Nealon} R.,   {Pinte} C.,  2020,
  \mn@doi [\apjl] {10.3847/2041-8213/abc704}, \href
  {https://ui.adsabs.harvard.edu/abs/2020ApJ...904L..18R} {904, L18}

\bibitem[\protect\citeauthoryear{{Rowther}, {Nealon}  \& {Meru}}{{Rowther}
  et~al.}{2022}]{2022Rowther}
{Rowther} S.,  {Nealon} R.,   {Meru} F.,  2022, \mn@doi [\apj]
  {10.3847/1538-4357/ac3975}, \href
  {https://ui.adsabs.harvard.edu/abs/2022ApJ...925..163R} {925, 163}

\bibitem[\protect\citeauthoryear{{Rowther}, {Nealon}  \& {Meru}}{{Rowther}
  et~al.}{2023}]{2022bRowther}
{Rowther} S.,  {Nealon} R.,   {Meru} F.,  2023, \mn@doi [\mnras]
  {10.1093/mnras/stac3106}, \href
  {https://ui.adsabs.harvard.edu/abs/2023MNRAS.518..763R} {518, 763}

\bibitem[\protect\citeauthoryear{Segura-Cox et~al.,}{Segura-Cox
  et~al.}{2020}]{Segura-Cox2020}
Segura-Cox D.~M.,  et~al., 2020, \mn@doi [Nature] {10.1038/s41586-020-2779-6},
  586, 228

\bibitem[\protect\citeauthoryear{{Sheehan} \& {Eisner}}{{Sheehan} \&
  {Eisner}}{2018}]{2018Sheehan}
{Sheehan} P.~D.,  {Eisner} J.~A.,  2018, \mn@doi [\apj]
  {10.3847/1538-4357/aaae65}, \href
  {https://ui.adsabs.harvard.edu/abs/2018ApJ...857...18S} {857, 18}

\bibitem[\protect\citeauthoryear{{Shi}, {Zhu}, {Stone}  \& {Chiang}}{{Shi}
  et~al.}{2016}]{2016Shi}
{Shi} J.-M.,  {Zhu} Z.,  {Stone} J.~M.,   {Chiang} E.,  2016, \mn@doi [\mnras]
  {10.1093/mnras/stw692}, \href
  {https://ui.adsabs.harvard.edu/abs/2016MNRAS.459..982S} {459, 982}

\bibitem[\protect\citeauthoryear{{Stamatellos} \& {Inutsuka}}{{Stamatellos} \&
  {Inutsuka}}{2018}]{2018Stamatellos}
{Stamatellos} D.,  {Inutsuka} S.-i.,  2018, \mn@doi [\mnras]
  {10.1093/mnras/sty827}, \href
  {https://ui.adsabs.harvard.edu/abs/2018MNRAS.477.3110S} {477, 3110}

\bibitem[\protect\citeauthoryear{{Toomre}}{{Toomre}}{1964}]{1964Toomre}
{Toomre} A.,  1964, \mn@doi [\apj] {10.1086/147861}, \href
  {https://ui.adsabs.harvard.edu/abs/1964ApJ...139.1217T} {139, 1217}

\bibitem[\protect\citeauthoryear{{Walmswell}, {Clarke}  \&
  {Cossins}}{{Walmswell} et~al.}{2013}]{2013Walmswell}
{Walmswell} J.,  {Clarke} C.,   {Cossins} P.,  2013, \mn@doi [\mnras]
  {10.1093/mnras/stt314}, \href
  {https://ui.adsabs.harvard.edu/abs/2013MNRAS.431.1903W} {431, 1903}

\bibitem[\protect\citeauthoryear{{Weidenschilling}}{{Weidenschilling}}{1977}]{1977Weidenschilling}
{Weidenschilling} S.~J.,  1977, \mn@doi [\mnras] {10.1093/mnras/180.2.57},
  \href {https://ui.adsabs.harvard.edu/abs/1977MNRAS.180...57W} {180, 57}

\bibitem[\protect\citeauthoryear{{Wurster} \& {Bonnell}}{{Wurster} \&
  {Bonnell}}{2023}]{2023Wurster}
{Wurster} J.,  {Bonnell} I.~A.,  2023, \mn@doi [\mnras]
  {10.1093/mnras/stad1022}, \href
  {https://ui.adsabs.harvard.edu/abs/2023MNRAS.522..891W} {522, 891}

\makeatother
\end{thebibliography}




\appendix

\section{Clump Finding}

\begin{figure}
    \centering
    \includegraphics[width=\linewidth]{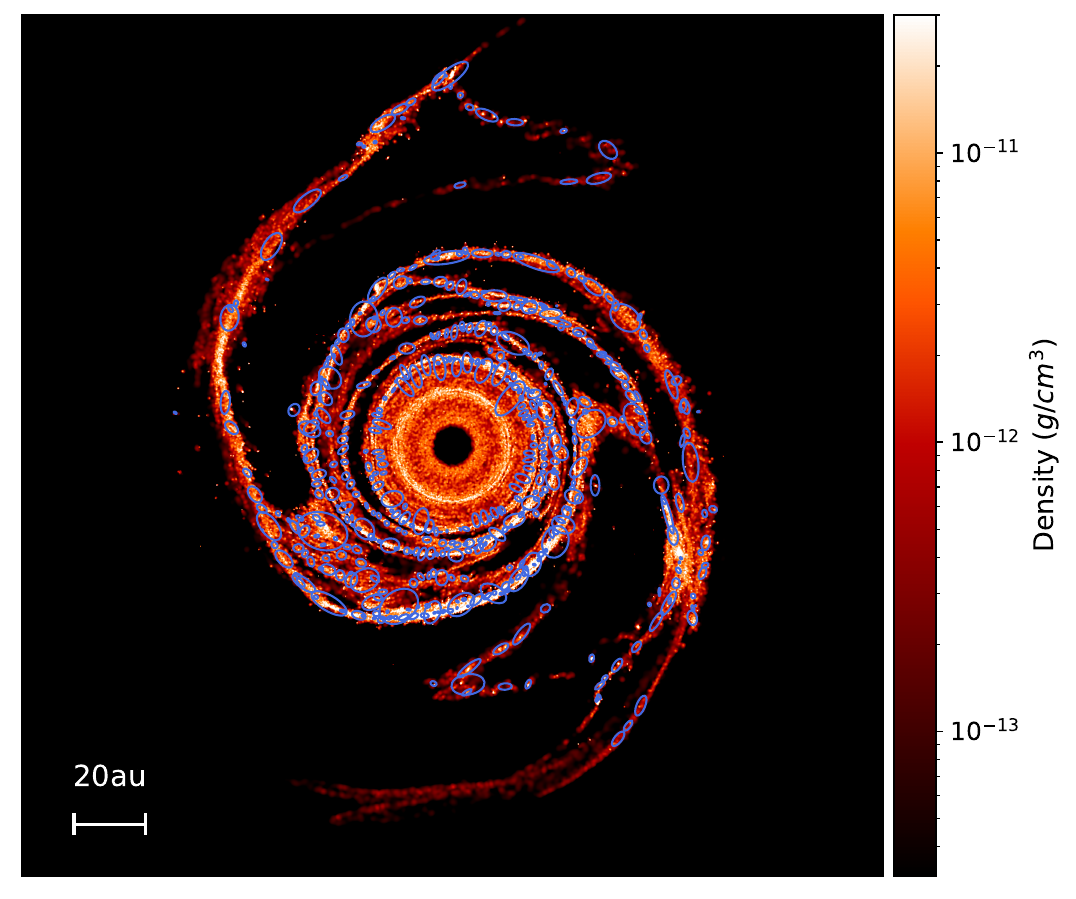}
    \caption{Locations of the clumps detected using the clump finding algorithm (\S\ref{sec:clumps}) plotted on top of a cross section slice of the dust density in the $z=0$ plane.}
    \label{fig:clumps}
\end{figure}

Figure \ref{fig:clumps} plots the locations of the clumps detected (blue ellipses) using the clump finding algorithm described in \S\ref{sec:clumps} on top of the dust density for the simulation with 50\,cm sized grains. In general, the clumps are detected within the gas spirals where dust is trapped.  The clumps do not always appear to lie on top of bright (dense) spots due to the clumps being detected in 3D. Whereas the density is a 2D projection. Additionally, clumps that have less than 58 neighbours are considered unresolved and are ignored.

\begin{figure*}
    \centering
    \includegraphics[width=\linewidth]{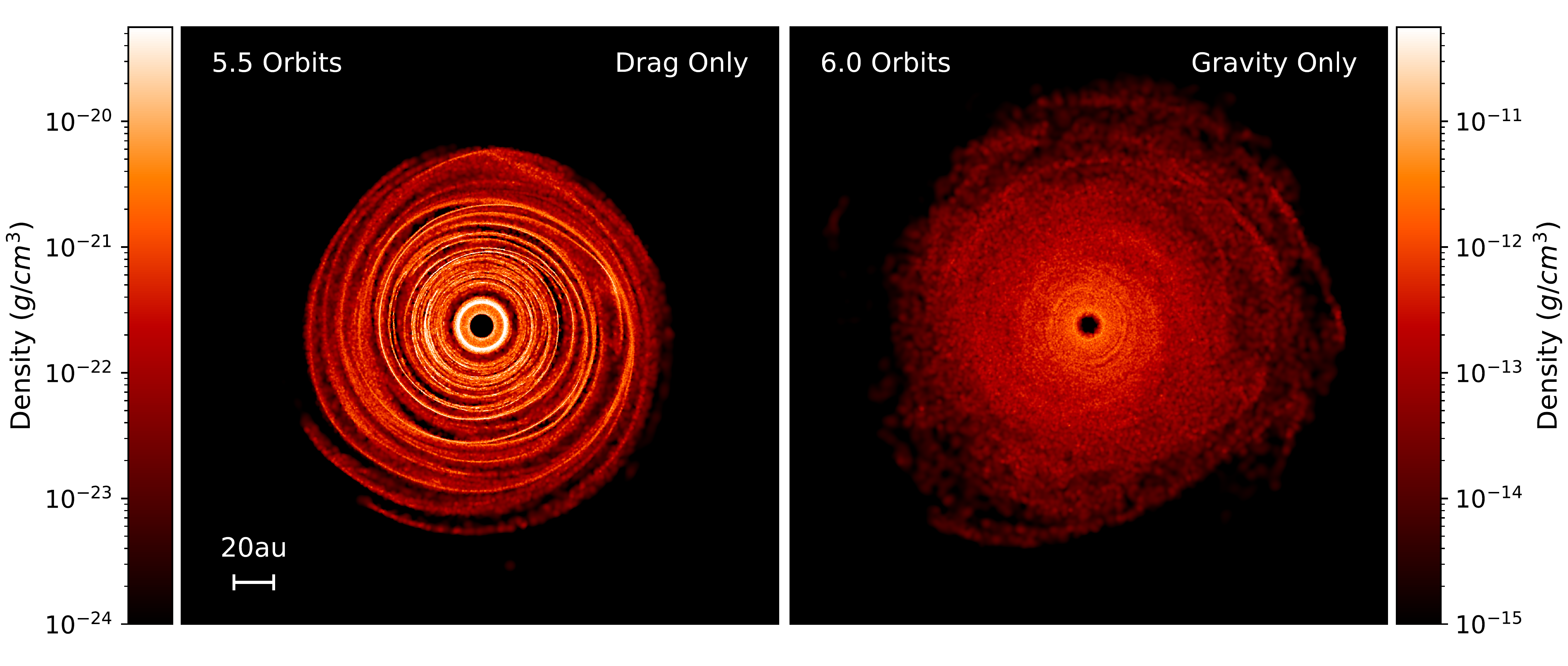}
    \caption{Cross section slices of dust density in the $z=0$ plane showing the evolution of the bottom two panels in Figure \ref{fig:allPhysics}. The two simulations are where the dust no longer feels gravitational acceleration from either the gas or dust (Drag Only - left panel), and the dust no longer feels the drag force (Gravity Only - right panel). The spirals that were present in Fig \ref{fig:allPhysics} eventually disappear if the dust only feels drag or gravity. The combined influence of drag and gravity is required for dust to remain trapped in spirals.}
    \label{fig:longTerm}
\end{figure*} 

\section{Long(er) term evolution}
\label{app:longTerm}

Figure \ref{fig:longTerm} shows the evolution of the Drag Only and Gravity Only simulations. From Figure \ref{fig:allPhysics}, it initially appears that the dust is able form spiral structures solely due to either drag or dust. However, as seen from Fig \ref{fig:longTerm} the spirals disappear soon after, highlighting that for dust to stay trapped in spirals, both drag are gravity are necessary.



\bsp	
\label{lastpage}
\end{document}